\documentclass[11pt]{article}
\usepackage[left=2cm, right=2cm, top=2cm]{geometry}

\usepackage[english]{babel}

\usepackage{pdfpages}

\usepackage{algpseudocode,algorithm}
\usepackage{listings}
\usepackage{amsmath}
\usepackage{qtree}
\usepackage{graphicx}
\usepackage{caption}
\usepackage{subcaption}
\usepackage{hhline}
\usepackage{xcolor}
\usepackage{tabularx} 
\usepackage{collcell}
\usepackage{multirow}
\usepackage{booktabs}

\usepackage{setspace}

\usepackage{adjustbox}

\usepackage{array}
\usepackage{makecell}

\date{}

\author{Willian de Oliveira Barreiros J\'unior and George Teodoro}%

\title{Accelerating Sensitivity Analysis in Microscopy Image Segmentation Workflows}%

\providecommand{\keywords}[1]
{
  \small
  \textbf{\textit{Keywords---}} #1
}

\begin{document}%
	\maketitle
	\setstretch{1.2}

	\begin{abstract}
	With the increasingly availability of digital microscopy imagery equipments there is a demand for efficient execution of whole slide tissue image applications. Through the process of sensitivity analysis it is possible to improve the output quality of such applications, and thus, improve the desired analysis quality. Due to the high computational cost of such analyses and the recurrent nature of executed tasks from sensitivity analysis methods (i.e., reexecution of tasks), the opportunity for computation reuse arises. By performing computation reuse we can optimize the run time of sensitivity analysis applications. This work focuses then on finding new ways to take advantage of computation reuse opportunities on multiple task abstraction levels. This is done by presenting the coarse-grain merging strategy and the new fine-grain merging algorithms, implemented on top of the Region Templates Framework.
	\end{abstract}

	\hspace{10pt}

	\hspace{10pt}

    \section{Introduction}

We define algorithm sensitivity analysis (SA) as the process of quantifying, comparing, and correlating output from multiple analyses of a dataset computed with variations of an analysis workflow using different input parameters \cite{sa}. This process is executed in many phases of scientific research and can be used to lower the effective computational cost of analysis on such researches, or even improve the quality of the results through parameter optimization.

The main motivation of this work is the use of image analysis workflows for whole slide tissue images analysis \cite{motiv1}, which extracts salient information from tissue images in the form of segmented objects (e.g., cells) as well as their shape and texture features. Imaging features computed by such workflows contain rich information that can be used to develop morphological models of the specimens under study to gain new insights into disease mechanisms and assess disease progression.

A concern with automated biomedical image analysis is that the output quality of an analysis workflow is highly sensitive to changes in the input parameters. As such, adaptation of SA methods and methodologies employed in other fields \cite{moat,vbd,motiv2,motiv3}, can help understanding image analysis workflows for both developers and users. In short, the benefits of SA include: (i) better assessment and understanding of the correlation between input parameters and analysis output; (ii) the ability to reduce the uncertainty / variation of the analysis output by identifying the causes of variation; and (iii) workflow simplification by fixing parameters values or removing parts of the code that have limited or negligible effect on the output.

Although the benefits of using SA are many, its use in practice is limited given the data and computation challenges associated with it. For instance, a single study using a classic method such as MOAT (Morris One-At-Time) \cite{moat} may require hundreds of runs of the image analysis workflow (sample size). The execution of a single Whole Slide Tissue Image (WSI) will extract about 400,000 nuclei on average and can take hours on a single computing node. A study at scale will consider hundreds of WSIs and compute millions of nuclei per run, which need to be compared to a reference dataset of objects to assess and quantify differences as input parameters are varied by the SA method. A single analysis at this scale using a moderate sample size with 240 parameter sets and 100 WSI would take at least three years if executed sequentially \cite{rtf1}. Given how time consuming such analysis is, there is a demand to develop mechanisms to make it feasible, such as parallel execution of tasks and computation reuse.

The information generated with a SA method is computed by executing or evaluating the same workflow as values of the parameters are systematically varied. As such, there are several parameters sets which have parameters with similar values. The workflows used on this work are hierarchical and, as such, can be broken down in routines, or fine-grained tasks. As such, it would be wasteful if one of these routines were to be executed on two or more evaluations generated by the SA method with the same parameters values and inputs. Thus, the re-executions of a given routine could reuse the results of the first execution in order to reduce the overall cost of the application.

Formally, computation reuse is the process of reusing routines or tasks results instead of re-executing them. Computation reuse opportunities arise when multiple computation tasks have the same input parameters, resulting in the same output, and thus making the re-execution of such task unnecessary. Computation reuse can also be classified by the level of abstraction of the reused tasks. Furthermore, these tasks can be combined on hierarchical workflows, with the routines and sub-routines of which they are composed by, being able to be fully or partially reused. Seizing reuse opportunities is done by a merging process, in which two or more tasks are merged together, after which the repeated or reusable portions of the merged tasks are executed only once.

Computation reuse on this work will be accomplished with the use of finer-grain tasks merging algorithms, as opposed to the already existing coarser-grain merging method implemented on the  Region Templates Framework (RTF) platform, on which all algorithms are implemented on. This platform is responsible for the distributed execution of hierarchical workflows in large-scale computation environments.


Other works have studied computation reuse as a means to reduce overall computational cost in different ways \cite{reuse1,reuse2,reuse3,reuse4,reuse5,reuse6,reuse7,reuse8}. Although the principle of computation reuse is rather abstract, its implementation on this work is distinct from existing methods. Some of these methods resort to hardware implementations \cite{reuse2,reuse3}, which are not general or flexible enough for the given problem. Some apply reuse by profiling the application \cite{reuse4}, which is also impracticable on the SA domain. Finally, most of them rely on caching systems of distinct levels of abstraction to reduce the overall cost of the applications \cite{reuse5,reuse6,reuse7,reuse8}, being too expensive to employ on the desired scale of distribution. 


Summarizing, this work focuses on two ways of accomplishing computation reuse in SA applications for the RTF: (i) coarse-grain tasks reuse and (ii) fine-grain tasks reuse. The main differences between them, besides the granularity of the tasks to be reused, are the underlying restrictions of the system used to execute these tasks. The reuse of coarse-grain tasks can offer a greater speedup when reuse happens, but there are less reuse opportunities. With fine-grain tasks these reuse opportunities are more frequent, however, more sophisticated strategies need to be employed in order to deal with dependency resolution and to avoid performance degradation due to the impact of excessive reuse to the parallelism.


\subsection{The Problem}

Because of high computing demands, sensitivity analysis applied to microscopy image analysis is unfeasible for routinely use when applied to whole slide tissue images.

\subsection{Contributions}

This work focuses on improving the performance of SA studies in microscopy image analysis through the application of finer-grain computation reuse on top of the coarse-grain computation reuse.



The specific contributions of this work are presented below with a reference to the section in which they are described:

\begin{enumerate}
  \item A graphical user interface for simplifying the deployment of workflows for the RTF, which is coupled with a code generator that allows the flexible use of the RTF on distinct domains [Section \ref{sec:improve}];
  
  \item The development and analysis of multi-level reuse algorithms:
  \begin{enumerate}
    \item A coarse-grain merging algorithm was implemented [Section \ref{sec:stage-merging}];
    \item A fine-grain Na\"ive Merging Algorithm was proposed and implemented [Section \ref{sec:naive-merging}];
    \item The fine-grain Smart Cut Merging Algorithm was proposed and implemented [Section \ref{sec:sca}];
    \item The fine-grain Reuse-Tree Merging Algorithm was proposed and implemented [Section \ref{sec:rtma}];
  \end{enumerate}
  
  \item Proposal and implementation of the Task-Balanced Reuse-Tree Merging Algorithm that reduces the issue of loss of parallelism due to load imbalance provoked by the Reuse-Tree Merging Algorithm [Section \ref{sec:TRTMA}];
  
  \item The performance gains of the proposed algorithms with a real-world microscopy image analysis application were demonstrated using different SA strategies (e.g MOAT and VBD) at different scales.

  
\end{enumerate}

The contributions of Sections \ref{sec:improve}, \ref{sec:stage-merging} and \ref{sec:rtma} were published on the IEEE Cluster 2017 conference \cite{barreiros_parallel_2017}, comprising, but not restricted to the proposal of multi-level computational reuse, the proposal of the Reuse-Tree structure for fine-grain merging and the experimental results for lower scale tests. Moreover, the contributions of Section \ref{sec:TRTMA} are currently being drafted for a submission for a journal publication. These contributions, which are an extension of the published work \cite{barreiros_parallel_2017}, includes some further analysis of computation reuse within the scope of scalability under more challenging settings, some limitations of the previously proposed solution for fine-grain merging, and finally, a new approach to cope with such limitations.

\subsection{Document Organization}

The next section describes the motivating application, the theory behind computation reuse and the Region Templates Framework (RTF), which was used to deploy the application on a parallel machine and is also the tool in which the merging algorithms were incorporated. After these considerations more relevant related work is analyzed. Section III describes the proposed solutions for multi-level computation reuse, their implementations and optimizations. On Section IV the experimental procedures are described and the results are analyzed. Finally, Section V closes this work with contributions and possible future goals for its continuation.

    \section{Background}

This chapter describes the motivating application along with the Region Templates Framework, in which this work is developed, and some basic concepts of sensitivity analysis and computation reuse. Being the contributions of this work restricted to computation reuse, this chapter then ends with the analysis of some relevant related work on the subject.

\subsection{Microscopy Image Analysis}

It is now possible for biomedical researchers to capture highly detailed images from whole slide tissue samples in a few minutes with high-end microscopy scanners, which are becoming evermore available. This capability of collecting thousands of images on a daily basis extends the possibilities for generating detailed databases of several diseases. Through the investigation of tissue morphology of whole slide tissue images (WSI) there is the possibility of better understanding disease subtypes and feature distributions, enabling the creation of novel methods for classification of diseases. With the increasing number of research groups working and developing richer methods for carrying out quantitative microscopy image analyses \cite{micro1, micro2, micro3, micro4, micro5, micro6, micro7, micro8} and also the increasingly availability of digital microscopy imagery equipment, there is a high demand for systems or frameworks oriented towards the efficient execution of microscopy image analysis workflows.


\begin{figure}[b]
\begin{center}
\includegraphics[width=1\textwidth]{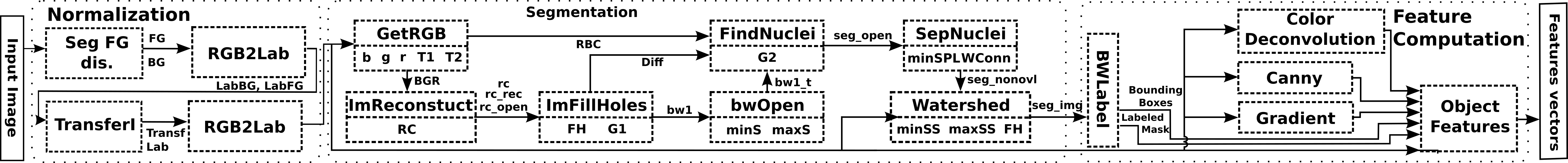}
\caption{An example microscopy image analysis workflow performed before image classification. Image extracted from \cite{barreiros_parallel_2017}.}
\label{fig:wf}
\end{center}
\end{figure}

\begin{figure}[t]
   \centering
   \begin{subfigure}[b]{0.45\textwidth}
       \includegraphics[width=\textwidth]{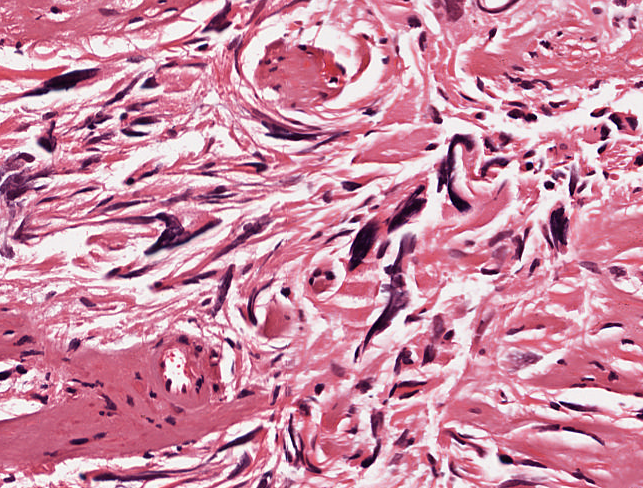}
       \caption{A tissue image.}
       \label{fig:rg1}
   \end{subfigure}
   \hspace{3mm}
   \begin{subfigure}[b]{0.45\textwidth}
        \includegraphics[width=\textwidth]{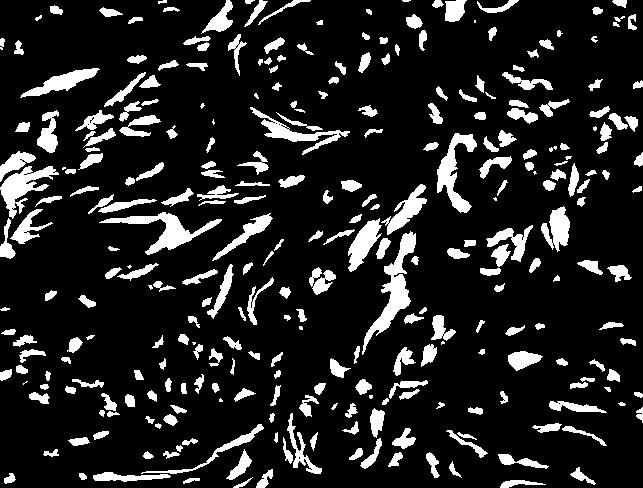}
        \caption{The segmented tissue image.}
        \label{fig:rg2}
    \end{subfigure}
    \caption{An example of tissue image segmentation.}
    \label{fig:tissue}
\end{figure}

The microscopy image analysis workflow used on this work is presented in Figure \ref{fig:wf} and was proposed by \cite{kong2013machine,DBLP:conf/bibm/KongWTCMKPSB13,Teodoro-IPDPS2013,6267914,Kong2015AutomatedCS,taveira,8512961}. This workflow consists of normalization (1), segmentation (2), feature computation (3) and final classification (4), being the first three analysis stages the most computationally expensive phases. The first stage is responsible for normalizing the staining and/or illumination conditions of the image. The segmentation is the process of identifying the nucleus of each cell of the analyzed image (Figure \ref{fig:tissue}). Through feature computation a set of shape and texture features is generated for each segmented nucleus. At last, the final classification will typically involve using data mining algorithms on aggregated information, by which some insights on the underlying biological mechanism that enables the distinction of subtypes of diseases are gained.

The quality of the workflow analysis is, however, dependent of the quality of the parameters values, with them described in Table \ref{tab:parameters}. Therefore, in order to improve the effectiveness of the analysis the impact of these parameters on the output of the used workflow (Figure \ref{fig:wf}) should be analyzed. This impact analysis is known as sensitivity analysis and is detailed on the following section.

\subsection{Sensitivity Analysis}


We define Sensitivity Analysis (SA) as the process of quantifying, comparing and correlating the input parameters of a workflow with the intent of quantifying the impact of each input to the final output of the workflow \cite{sa}. This process is applied on several phases of scientific research including, but not limited to model validation, parameter studies and optimization, and error estimation \cite{sa2}. The outcome of such methods, as defined in \cite{rtf2}, are statistics that quantify variance in the analysis results as well as measures such as sensitivity induces that indicate the amount of variance in the analysis results that can be attributed to individual parameters or combinations of parameters.

Usually, the computational cost for performing SA on a workflow is directly proportional to the number of parameters it has. One way to simplify the analysis on applications with large numbers of parameters, thus reducing its cost, is through the removal of parameters whose effect on the output is negligible.

This work focuses on using the already existing system, the Region Templates Framework (RTF) \cite{rtf1,rtf2}, which performs sensitivity analysis in two phases. On the first phase the 15 input parameters (Table \ref{tab:parameters}) are screened with a light, or less compute demanding,  SA method, used to remove the so called non-influential parameters from the next phase. Afterwards, a second SA method is executed on the remaining parameters, on which both first-order and high-order effects of these on the application output are quantified. This two-phase analysis is performed since the cost of more specific approaches (e.g., VBD) are prohibitively expensive.

\begin{table}[t!]
\begin{center}
\begin{scriptsize}
\begin{tabular}{l l l l }
\hline
Parameter   & Description & Range Values      \\ \hline
B/G/R & Background detection thresholds &  $[210,220,...,240]$  \\ \hline
T1/T2  & Red blood cell thresholds & $[2.5,3.0,...,7.5]$   \\ \hline
\multirow{2}{*}{G1/G2}  & Thresholds to identify  & $[5,10,...,80]$   \\ 
   & candidate nuclei & $[2,4,...,40]$    \\ \hline
MinSize(minS)   & Candidate nuclei area threshold   & $[2,4,...,40]$    \\ \hline
MaxSize(maxS)   & Candidate nuclei area threshold & $[900,..,1500]$   \\ \hline 
MinSizePl \\ (minSPL) & Area threshold before watershed & $[5,10,...,80]$   \\ \hline
MinSizeSeg \\ (maxSS) & Area threshold in final output   & $[2,4,...,40]$    \\ \hline 
MaxSizeSeg \\ (minSS) & Area threshold in final output  & $[900,..,1500]$   \\ \hline 
FillHoles(FH)  & propagation neighborhood  & $[4$-conn$,8$-conn$]$   \\ \hline
MorphRecon(RC) & propagation neighborhood  & $[4$-conn$,8$-conn$]$   \\ \hline
Watershed(WConn)  & propagation neighborhood  & $[4$-conn$,8$-conn$]$   \\ \hline
\end{tabular}
\end{scriptsize}
\caption{Definition of parameters and range values: parameter space contains about 21 trillion points.\label{tab:parameters}}
\end{center}
\vspace{-4mm}
\end{table}

This multi-phase sensitivity analysis process is approached on \cite{sa2} as an alternative to cope with costly analysis. The application case, as seen in Figure \ref{fig:wf} uses a complex model with several input parameters (see Table \ref{tab:parameters}) and a high execution cost. As such, it is recommended that a lighter preliminary analysis method should be executed on the full range of input parameters, only to reduce these to a smaller subset of important parameters. As a way to further reduce the analysis complexity on this first screening analysis is to also drop inputs' correlation analysis. After the execution of a screening method, more complex and comprehensive analysis methods can be performed on a subset of the input parameters. The chosen SA methods for this work were Morris One-At-A-Time as a screening method \cite{moat}, and Variance-Based Decomposition as a more complete analysis.

The light SA method, Morris One-At-A-Time (MOAT) \cite{moat}, performs a series of runs of the application changing each parameter individually, while fixing the remaining parameters in a discretized parameter search space. Each of the $k$ analyzed parameters values ranges are uniformly partitioned in $p$ levels, thus resulting in a $p^k$ grid of parameter sets to be evaluated. Each evaluation output $x_i$ of the application creates a parameter elementary effect (EE), calculated as $EE_i = \frac{y(x_1, ..., x_i + \Delta_i, ..., x_k)-y(x)}{\Delta_i}$, with $y(x)$ being the application output before the parameter perturbation. In order to account for global SA the RTF uses $\Delta_i = \frac{p}{2(p-1)}$ \cite{rtf2}. The MOAT method requires $r(k+1)$ evaluations, with $r$ in the range of 5 to 15 \cite{moat2}.

The second SA method, Variance-Based Decomposition (VBD) is preferably performed after a lighter SA screening method, as the MOAT method. This is done since VBD requires $n(k+2)$ evaluations for $k$ parameters and $n$ samples, with $n$ lying in the order of thousands of executions \cite{vbd}. Thus, it is interesting to use a reduced number of parameters for feasibility reasons. VBD, unlike MOAT, discriminates the the output uncertainty effects among individual parameters (first-order) and high-order effects.

\begin{table}[]
\begin{center}
\begin{scriptsize}
\begin{tabular}{lc|lccc}
\hline
MOAT       & First-order Effect & VBD        & First-order Effect (Main) & Higher-order Effects (Total) \\
\hline
B          & -0,0108            & B          & -                  & -                    \\
G          & -0,0064            & G          & -                  & -                    \\
R          & -0,0189            & R          & -                  & -                    \\
T1         & 0,0207             & T1         & -                  & -                    \\
T2         & 0,0417             & T2         & 0,0006             & 0,0001               \\
G1         & 0,8157             & G1         & 0,2251             & 0,2371               \\
G2         & 0,9197             & G2         & 0,7305             & 0,7886               \\
MinSize    & 0,0889             & MinSize    & 0,0025             & 0,0056               \\
MaxSize    & 0,1820             & MaxSize    & 0,0150             & 0,0086               \\
MinSizePl  & 0,0341             & MinSizePl  & 0,0021             & 0,0022               \\
MinSizeSeg & -0,0155            & MinSizeSeg & -                  & -                    \\
MaxSizeSeg & -0,0184            & MaxSizeSeg & -                  & -                    \\
FillHoles  & -0,0276            & FillHoles  & -                  & -                    \\
MorphRecon & 0,1321             & MorphRecon & 0,0146             & 0,0129               \\
Watershed  & 0,0530             & Watershed  & 0,0018             & 0,0016               \\
\hline
\end{tabular}
\end{scriptsize}
\caption{Example output of a MOAT analysis with all 15 parameters and a VBD analysis with a selection of the 8 most influential parameters. The influence of a parameter is bounded in the interval [-1,1] and is proportional to its distance from 0 (i.e., 1 and -1 are the greatest values and 0 the smallest).}
\label{tab:sa-out-example}
\end{center}
\end{table}

As an example, Table \ref{tab:sa-out-example} provides the expected outcome of a two-steps SA of the used workflow. The first analysis, MOAT, is performed at an earlier moment in order to screen all parameters regarding their first-order effects or influence over the output. Afterwards, the VBD analysis can be performed with a subset of the 8 most influential parameters, yielding not only more precise first-order effect values but also a way to calculate higher-order effects through the manipulation of the {\it Total} values (e.g., for a third-order effect of T2, G1 and G2, their {\it Total} values are added together and compared with the remaining {\it Total} values).

Regardless of the SA method chosen, the use of large set of parameters (Table \ref{tab:parameters}) results in the unpractical task of performing SA on the workflow of Figure \ref{fig:wf} due to the expected cost of evaluating such large search domain. For the sake of mitigating this infeasibility issue for performing SA on the presented workflow we can execute the analysis on high-end distributed computing environments. Also, computation reuse can be employed to reduce the computational cost without the need of application specific optimizations. Both  mentioned methods are described in the next sections.

\subsection{Region Templates Framework (RTF)}

\begin{figure}[b!]
\begin{center}
\includegraphics[width=0.95\textwidth]{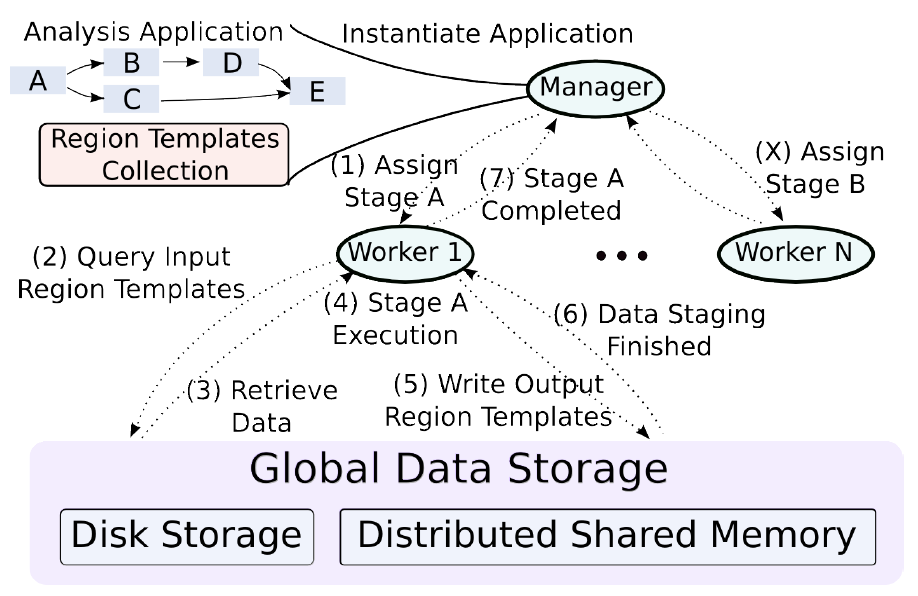}
\caption{The main components of the Region Templates Framework, highlighting the steps of a coarse-grain stage instance execution. Image extracted from \cite{rtf1}.}
\label{fig:ss}
\end{center}
\end{figure}

The Region Template Framework (RTF) abstracts the execution of a workflow application on distributed environments \cite{rtf1}. It supports hierarchical workflows that are composed of coarse-grain stages, which in turn are composed by fine-grain tasks. The dependencies between stages, and tasks of a single stage are solved by the RTF. Given a homogeneous environment of $n$ nodes with $k$ cores each, any stage instance must be executed on a single node, with its tasks being executed on any of the $k$ cores of the same node. It is noteworthy that, not only any node can have more than one stage instance executing on it, but also, there may be  more than one task from the same stage running in parallel, given that the inter-tasks dependencies are respected.

\begin{figure}[t!]
\begin{center}
\includegraphics[width=0.95\textwidth]{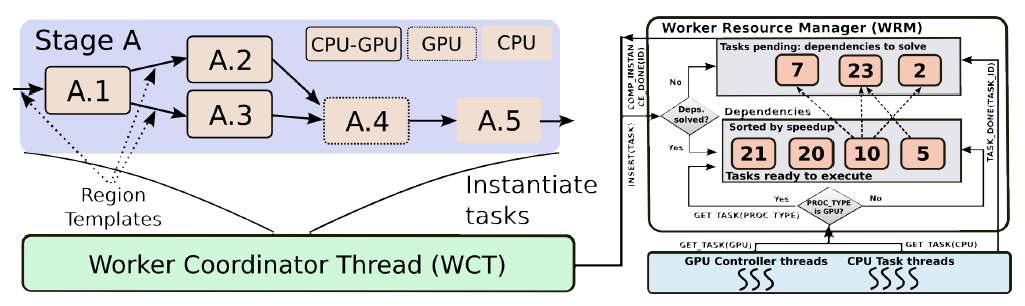}
\caption{The execution of a stage instance from the perspective of a node, showing the fine-grain tasks scheduling. Image extracted from \cite{rtf1}.}
\label{fig:ts}
\end{center}
\vspace{-4mm}
\end{figure}

The main components of the RTF are: the data abstraction, the runtime system, and the hierarchical data storage layer \cite{rtf1}. The runtime system consists of core functions for scheduling of application stages, transparent data movement and management via the storage layers. Figure \ref{fig:ss} shows an example of the dispatch of a stage to a worker with the data exchanges in the RTF storage layer. The RTF, with its centralized Manager, distributes the stages to be executed to Worker nodes across the network. The hierarchical workflow representation allows for different scheduling strategies to be used at each level (stage-level and task-level). Fine-grain scheduling is possible at task-level in order to also exploit variability in performance of application operations in hybrid systems. In Figure \ref{fig:ts} a stage A is sent to a worker node for execution, which tasks are scheduled locally. 

Still on the scheduler, the Manager schedules stages to Workers on a
demand-driven basis, with the Workers requesting work from the Manager until
all stages are executed. Since the Worker decides when they request more work,
a Worker can execute one or more stage at any given time instant, based on its
underlying infrastructure. Being a stage composed of tasks, these are scheduled
locally by the Worker executing them. These tasks differ in terms of data
access patterns and computation intensity, thus, attaining different speedups
if executed on co-processors or accelerators. In order to optimize the
execution of tasks a Performance Aware Task Scheduling (PATS) was implemented
\cite{rtf1,Teodoro-IPDPS2013,Teodoro:2014:CPA:2650283.2650645,cluster09george,CPE:CPE4425,doi:10.1177/1094342015594519,DBLP:journals/pc/TeodoroPKKCS13}.
With PATS, tasks are assigned to either a CPU or GPU core based on its
estimated acceleration and the current device load.

On the data storage layer the Region Templates (RT) data abstraction is used to represent and interchange data (represented by the collection of objects of an application instance and the stored data of Figure \ref{fig:ss}). It consists of storage containers for data structures commonly found in applications that process data in low-dimensional spaces (1D, 2D or 3D spaces) with a temporal component. The data types include: pixels, points, arrays (e.g., images or 3D volumes), segmented and annotated objects and regions, all of which are implemented using the OpenCV \cite{opencv} library interfaces to simplify their use. A RT data instance represents a container for a region defined by a spatial and temporal bounding box. A data region object is a storage materialization of data types and stores the data elements in the region contained by a RT instance, which may have multiple data regions. 

Access to the data elements in data regions is performed through a lightweight class that encapsulates the data layout, provided by the RT library. Each data region of one or multiple RT instances can be associated with different data storage implementations, defined by the application designer. With this design  the decisions regarding data movement and placement are delegated to the runtime environment, which may use different layers of a system memory to place the data according to the workflow requirements.



The runtime system is implemented through a Manager-Worker execution model that combines a bag-of-tasks execution with workflows. The application Manager creates instances of coarse-grain stages, and exports the dependencies among them. These dependencies are represented as data regions to be consumed/produced by the stages. The assignment of work from the Manager to Worker nodes is performed at the granularity of a stage instance using a demand-driven mechanism, on which each Worker node independently requests stages instances from the Manager whenever it has idle resources. Each node is then responsible for fine-grain task scheduling of the received stage(s) to its local resources.

To create an application for the RTF the developer needs to provide a library of domain specific data analysis operations (in this case, microscopy image analysis) and implement a simple startup component that generates the desired workflow and starts the execution. The application developer also needs to specify a partitioning strategy for data regions encapsulated by the region templates to support parallel computation of said data regions associated with the respective region templates.

Stages of a workflow consume and produce Region Template (RT) objects, which are handled by the RTF, instead of having to read/write data directly from/to stages or disk. While the interactions between coarse-grain stages are handled by the RTF, the task of writing more complex, fine-grained, stages containing several external, domain specific, fine-grain API  calls is significantly harder for application experts. This occurs since the RTF works only with one type of task objects as its runnable interface, not providing an easy way to compose stages using fine-grain tasks. The RTF also supports efficient execution on hybrid systems equipped with CPU and accelerators (e.g, GPUs).

\subsection{Related Work on Computation Reuse}
\label{sec:reuse_intro}

The idea of work reuse, also known as value locality \cite{reuse10,reuse14}, has been employed on both the hardware and software fronts with diverse techniques, such as value prediction \cite{reuse1}, dynamic instruction reuse \cite{reuse2} and $memoization$ \cite{reuse3}, with the goal of accelerating applications through the removal of duplicated computational tasks. This concept has been used for runtime optimizations on embedded systems \cite{reuse4}, low-level encryption value generation \cite{reuse5} and even stadium designing \cite{reuse6}. In order to further analyze these existing approaches as to find desirable features that solve the problem approached in this work a qualitative analysis was performed, which is summarized on Table \ref{tab:tax}. This analysis uses taxonomic terms defined here to classify computation reuse approaches. The proposed taxonomic terminology is explained on the next section in addition to a brief analysis of each of the studied computation reuse approaches.

\subsubsection{Computation Reuse Taxonomy}

\paragraph{Implementation Level (IL)}

Computation reuse can be enforced on either {\it Software} (S) or {\it Hardware} (H) levels. By {\it Software-Level} it is meant a hardware-independent approach that can either be executed as a static analysis before the execution of any computational task, or as a runtime approach that performs computation reuse as the application is executed. Also, it is possible for computation reuse to be searched on compilation-time by a customized compiler, which is also defined as {\it Software-Level}. It is also possible for these techniques to be combined.


\paragraph{Application Flexibility}

Here we define the {\it Application Flexibility}(AF) of an approach as either {\it General}, {\it Partial} or {\it Domain Specific} (DS). A {\it General} approach is any that does not have domain-specific restrictions that limits or prevents its use on different domains. The flexibility of an approach can also be {\it Partial}, meaning that either some non-trivial adaptations need to be employed or that anything outside its application domain will execute rather poorly. If an approach can only be used on a rather specific environment, or under strict restrictions it is said to be a {\it Domain Specific} approach.

\paragraph{Reuse Strategy}

One of the most important computation reuse characteristics is how computation reuse opportunities are found and explored. These can be defined as {\it Predictive}, {\it Memoization} or {\it Analytic} approaches. Computation reuse can be attained through the speculative technique of {\it Value Prediction} \cite{reuse2,reuse10} with its implementations relying on a buffer that contains the results of previous instructions executions, with which the value prediction is performed.

The most common technique for computation reuse is through {\it Memoization}, which is a cache-based approach on which reusable tasks results are stored on a buffer for later reuse. It is worth noting that the stored values are used as-is, unlike with {\it Value Prediction}, which relies on the evaluation of the buffered values in order to return a reusable value. This approach has the drawback of needing a buffer structure, which increases the complexity of this kind of solution.



The alternative to {\it Memoization} is to find all reuse opportunities in an {\it Analytic} manner. This means that the reused tasks were found {\it a priori}, instead of searching the results in a buffer as with the {\it Memoization} scheme. While this approach is considered to be the one with the least overhead, such analysis is more difficult to be achieved.


\paragraph{Tasks Granularity}

Still another rather important aspect of computation reuse is the {\it Granularity} of the reusable tasks. On this work we break {\it Task Granularity} in four categories: {\it Instruction-Level} (i.e., CPU instruction), {\it Fine-Grain Subroutines}, {\it Coarse-Grain Routines} and {\it Full Application}. We differentiate {\it Fine-Grain} from {\it Coarse-Grain} tasks by their semantical meaning, and as a consequence, their overall cost. If a task is big enough to have a broader meaning (e.g., a segmentation operation) we call it a {\it Coarse-Grain Routine}. If the task is bigger than a CPU instruction but also not big enough to have a more abstract meaning (e.g., the preparation of a matrix on memory, or a set of loops on an algorithm) we define them as {\it Fine-Grain Subroutines} or tasks. Finally, some approaches may only be able to work with a {\it Full Application} execution.

The importance of the granularity for computation reuse is that it limits the maximum amount of reuse of any application. As an example we have a segmentation algorithm. If we were to break it in CPU instructions and then perform a complete search for reusability (i.e., search for all available reuse) we would attain the maximum possible reuse. However, the potential overhead for exploiting this level of reuse is high. By grouping this low-level operations into subroutines we reduce the number of tasks, making the search for reuse more feasible. This grouping would also hide some reuse opportunities, effectively reducing the reuse potential of the application. 


\paragraph{Reusable Tasks Matching (RTM)}

An easy way to improve the reuse degree of an application is by relaxating the matching constraint for reuse. By doing this, reuse is possible even if not all tasks' parameters match, unlike the most common case on which all tasks' parameters are the {\it Same}. The obvious consequence of doing this relaxation is that the tasks' results will be different. However, some applications can deal with small imprecisions of its tasks (e.g., neural networks, multimedia applications, floating-point operations). As such, given that these partial (or {\it Similar}) matchings respect the precision necessary for these applications which can cope with such imprecisions, this strategy can improve the amount of reuse available.



\paragraph{Reuse Evaluation}

Computation reuse can be analyzed either {\it Dynamically}, at runtime, of {\it Statically} before the execution of any task. 


\paragraph{Training Required (T)}

Approaches that rely on domain-specific characteristics of applications (e.g., neural networks) usually require a {\it Training} step before the reuse analysis. For these approaches it is important to be mindful of the {\it Training} cost.

\paragraph{Reusability Environment Scale (RES)}

The reusable tasks scope is defined here as the {\it Reusability Environment Scale}. The tasks can be reusable among a {\it Distributed} (D) environment of computing nodes or reused only {\it Locally} (L). 





\subsubsection{Related Work Analysis}

Sodani and Sohi \cite{reuse2} motivate their work by drawing a parallel of a computation \textit{reuse buffer} used to optimize instruction execution with memory cache used to optimize memory access instructions. Their approach aims to reduce computational cost through reuse by (i) ending the instruction pipeline earlier, thus also reducing resources conflicts, and (ii) by breaking dependencies of later instructions, which can be executed earlier since the necessary inputs are already present. They initially proposed their \textit{reuse buffer} as a way to reduce branch misprediction penalties. However, the effectiveness of this approach proved itself much more powerful since the reuse frequency of other, more generic, types of instructions also proved to be high. Their implementation focus on adding a \textit{reuse buffer} to any generic dynamically-scheduled superscalar processor, using one of the three instruction reuse schemes proposed by them. 

The approach on \cite{reuse2} can be used for any application domain while also being exposed to the largest possible amount of reuse opportunities. Their incorporation of the \textit{reuse buffer} in a superscalar processor is done without impacting the pipeline  critical path, thus having no negative impact on non-reused instructions. Nevertheless, the efficiency of all instruction reuse schemes are heavily reliant on the buffer size. Although the used buffer sizes tested by them are small, this dependency is a limiting factor for the approach since smaller buffers means less reuse opportunities. Finally, the use of a hardware-based approach limits its use even further given the difficulty to design a processor for this sole purpose.

The work on \cite{reuse3}, similarly to \cite{reuse2}, also uses hardware-level $memoization$, but this time with a subset of operations called \textit{trivial computation}. These are potentially complex operations that are trivialized by simple operands (e.g., integer division by two). This strategy greatly simplifies the reuse protocol (i.e., whether an instruction is reused, insertion and replacement policies) at the cost of reuse opportunities. The speedups achieved by this approach were only significant when the application was favorable to the reuse strategy (e.g., Ackerman-like applications with huge amounts of trivial operations, or floating-point-intensive programs, which have naturally long-latency instructions). The same limitations of \cite{reuse2} were present here as well.

Wang and Raghunathan \cite{reuse4} attempt to reduce the energetic cost of embedded software on battery-powered devices through a profiling-based reuse technique with a $memoization$ structure. Some interesting discussions risen in their work regard reusable tasks granularity and the limitations of hardware-based reuse. Hardware implementations of computation reuse are usually complex, and the use of overly fine-grained operations for reuse may yield little or negative speedups given the overhead of $memoization$ caches.

The methodology of \cite{reuse4} consists on profiling an application, generating \textit{computation reuse regions}, setting the software cache configuration, evaluating the energy expenditure and then doing it all over again until a good enough solution is found. Only then, the optimized application is sent to production. The concept of flexible \textit{computation reuse regions} is very powerful since it makes the application more domain-independent while also optimizing the granularity of the reuse for any application instance. Their automated software cache configuration is also interesting since any $memoization$-based technique is heavily reliant on its size and performance.

Unfortunately \cite{reuse4} do not specify the cost of profiling (since for the test environment the typical input traces of the selected benchmarks were already available), nor the cost of configuring the \textit{computation reuse regions} and the software cache. Regardless, this approach, while presenting the concepts of flexible granularity and automatic software cache configuration / optimization, cannot be recommended for large-scale workflow execution given its unknown-cost training step. Also, in order to distribute the computation reuse, the software cache used by it needs to be re-thought to be compatible with this paradigm.

It is brought to our attention on \cite{reuse5} the cost of two-party secure-function evaluation (SFE) and the tendency to offload these operations from resource-constrained devices to outside servers. In order to reduce the computational cost of these SFE operations as well as bandwidth requirements, a system on which state is retained as to later be reused was implemented. The reusable encrypted values can be used by a number of clients on a distributed setting, originating from a centralized server node that implements a $memoization$ buffer.

Although \cite{reuse5} is the first approach to enable computation reuse to be done in a distributed environment, the encrypted values buffer is a bottleneck for the approach scalability. In order to remove this bottleneck, the buffer can be distributed among server nodes, which has as a consequence either (i) the buffers are coherent, and as such the servers need to keep trading messages to enforce it, or (ii) the buffers are not coherent and thus the reuse potential is reduced. Finally, this approach is only partially applicable for different application domains since the granularity of the reused tasks must be rather coarse in order to achieve good speedups. This happens because the of the big overhead of reusing encrypted values.

Approach \cite{reuse7} also works with distributable reusable values, but this time with  bioinformatics applications, which are known to be computationally expensive. The granularity of reusable tasks is even coarser, being able to perform full end-to-end reuse of workflows. When comparing with \cite{reuse5}, \cite{reuse7} has the same limitations given its $memoization$-based approach.

On \cite{reuse8} Santos and Santos propose the use of a software-level runtime buffer system to cache and then reuse energy evaluations for predicting the native conformation of proteins. The domain-specific application relies on a genetic algorithm, and as such, their approach is tailored for this single application. A similar approach is the one of Yasoubi et al. \cite{reuse9}, regarding the use of $memoization$-based computation reuse, optimized for a specific domain, which is neural networks on this case.

Yasoubi et al. \cite{reuse9} propose an offload hardware accelerator that uses clustered groups of neurons that maximize the expected computation reuse when executing the application. It is worth noting that the clustering is done by a k-means algorithm on software level. The reusable tasks are hardware-level multiplication instructions that, given the multi-processing-unit (multi-PU) architecture, disable PUs that perform repetitive operations, thus reducing the power consumption.

The work of Connors and W.Hwu \cite{reuse10} exploit value locality through the combination of a hardware buffer, an instruction set for representing different-sized reusable computational tasks and a profile-guided compiler that groups instructions into reusable tasks as to optimize their granularity. This approach was implemented as a way to extend hardware-only-based reuse approaches while solving the limitation of instruction-level reusable tasks granularity. Again, the use of dynamically-sized reusable tasks makes the approach more flexible to different domains of applications while optimizing the reusable tasks granularity for each application instance. However, in order to implement this feature the approach on \cite{reuse10} limits itself by needing a complex hardware and compiler implementation and profiling information on the domain-specific application.

Álvarez et al. \cite{reuse11} focus on reducing the power consumption of low-end and/or mobile devices by applying computation reuse on multimedia applications. This is done by exploiting the imprecision tolerance of multimedia floating-point operations at hardware-level to reuse tasks that are similar enough, thus increasing the amount of attainable computation reuse. Nonetheless, this ``similar enough'' strategy limits the usability of this approach to multimedia applications, or applications which have a large number of floating-point reusable operations. The same is true for approach \cite{reuse13}, which in turn proposes a more generic implementation that was not tailored for multimedia applications.

The first analytic computation reuse method is presented by Xu et al., on \cite{reuse12}. On their work they propose a framework for Isogeometric Analysis (IGA) that reuse matrix calculations. The reuse operations were statically analyzed \textit{a priori} and are specific of IGA, meaning that this approach, although having good speedups, cannot be applied for other application domains.

Lepak and Lipasti \cite{reuse14} propose reuse of memory load instructions. This is done through the characterization of value locality for memory instructions and the implementations of two reuse protocols for both uniprocessed and multiprocessed environments. For uniprocessed systems reuse can be attained by either analyzing the value locality of specific instructions (based on the program structure), or the locality of a particular memory address (message-passing locality). Furthermore, they define \textit{silent stores} as stores operations that do not change the system state (i.e., the written value is the same as the one previously present on memory). Given some statistical analysis of how many \textit{silent stores} are on selected benchmarks, they set an ideal maximum reuse possible to be achieved and, through their proposed protocols, aim to get as close as possible to these values.

\begin{table}[t!]
\begin{center}
\vspace*{-2ex}
\begin{scriptsize}
\begin{tabular}{ccccccccc}
\toprule

Reference       & IL         & AF & Reuse Strat. & Tasks Granularity                                & RTM & Reuse Eval. & T & RES \\

\midrule

\cite{reuse1} & H            & Flexible                   & Predictive          & Instruction-Level                                & Same       & Dynamic          & No             & L   \\
\cite{reuse2} & H            & Flexible                   & Memoization         & Instruction-Level                                & Same       & Dynamic          & No             & L   \\
\cite{reuse3} & H            & Flexible                   & Memoization         & \makecell{Instruction-Level \\ Trivial Operations}             & Same       & Dynamic          & No             & L   \\
\cite{reuse4} & S            & Flexible                   & \makecell{Analytic + \\ Memoization}  & \makecell{Fine-Grain \\ Regions of Code} & Same       & Static           & Yes            & L   \\
\cite{reuse5} & S            & Partial                 & Memoization         & Coarse-Grain                          & Same       & Dynamic          & No             & D   \\
\cite{reuse7} & S            & Flexible                   & Memoization         & Full Application                                  & Same       & Dynamic          & No           & D  \\
\cite{reuse8} & S            & DS                    & Memoization         & Coarse-Grain          & Same       & Dynamic          & No             & L   \\
\cite{reuse9}       & H+S & DS                    & Memoization         & \makecell{Instruction-Level \\ Complex Operations} & Similar             & Dynamic          & Yes            & L   \\
\cite{reuse10}       & H+S & Partial        & Memoization         & \makecell{Instruction-Level + \\ Fine-Grain \\ Regions of Code} & Same       & \makecell{Static + \\ Dynamic}          & Yes     & L  \\
\cite{reuse11}       & H            & Partial & Memoization         & \makecell{Instruction-Level \\ Floating-Point \\ Operations}     & Similar             & Dynamic          & No             & L   \\
\cite{reuse12}       & S            & DS                    & Analytic            & Coarse-Grain                          & Same       & Dynamic          & No             & L         \\
\cite{reuse13}       & H            & Partial       & Memoization         & \makecell{Instruction-Level \\ Floating-Point \\ Operations}     & Similar             & Dynamic          & No             & L                 \\
\cite{reuse14}       & H            & Flexible                   & Memoization         & Instruction-Level                                & Same       & Static          & No             & L  \\
\midrule

Our Work       & S            & Partial                & Analytic            & \makecell{Coarse-Grain \\ and Fine-Grain}        & Same       & Static           & No             & D \\

\bottomrule

\end{tabular}
\end{scriptsize}
\caption{Taxonomic evaluation of computation reuse approaches. Implementation Level (IL): Hardware (H) or Software (S). Application Flexibility (AF): Flexible, Partial or Domain Specific (DS). Reuse Strategy: Predictive, Memoization or Analytic. Task Granularity: Instruction-Level, Fine-Grain, Coarse-grain or Full Application. Reusable Tasks Matching: Same or Similar. Reuse Evaluation: Static or Dynamic. Needs Training Step (T). Reusability Environment Scale (RES): Local (L) or Distributed(D).\label{tab:tax}}
\end{center}
\vspace{-4mm}
\end{table}

Since none of the previous applications is either compatible or flexible enough to work on the large scale bioinformatics workflows application domain, this work proposes a novel approach to computation reuse. The proposed approach works with software-level reuse, since it is being implemented on top of the RTF. Also, given that this application is supposed to be executed on a large-scale cluster environment, hardware-based approaches are impractical. Moreover, the runtime system must be light in order to execute on a large-scale distributed environment, thus making the use of $memoization$ impractical. Given that the application uses hierarchical workflows, any applications of other domains need to be converted to workflows in order to be executed by our approach, slightly impacting the application domain flexibility. Finally, computation reuse is achievable by a static analytic analysis of reuse before the execution of any task, thus removing any distribution limitations as long as the reuse analysis can be performed quickly.

    \section{Multi-Level Computation Reuse}

This work has as its main goal the development of Sensitivity Analysis (SA) optimizations through multi-level computation reuse. This chapter analyzes computation reuse and then describes improvements made to the Region Templates Framework (RTF), which were implemented in order to enable the use of multi-level computation reuse. After that, the new computation reuse approaches are described, along with their advantages and disadvantages.

\begin{figure}[t]
\begin{center}
\includegraphics[width=0.7\textwidth]{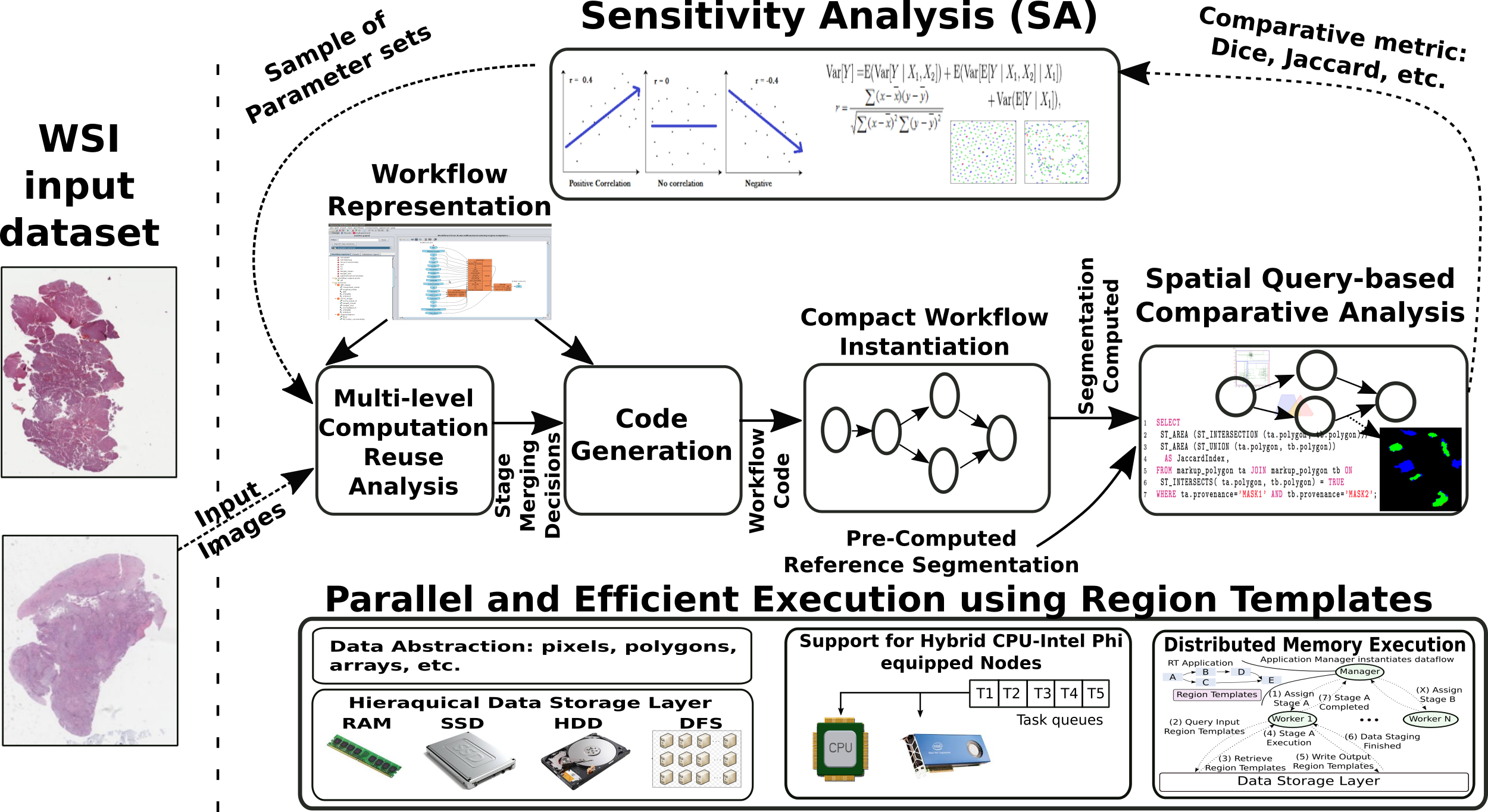} \caption{The parameter study framework. A SA method selects parameters of the analysis workflow, which is executed on a parallel machine. The workflow results are compared to a set of reference results to compute differences in the output. This process is repeated a set number of times (sample size) with varying input parameters' values.}
\label{fig:overview}
\end{center}
\vspace{-4mm}
\end{figure}

The SA studies and components that were developed and integrated into the RTF are illustrated in Figure \ref{fig:overview}. An SA study in this framework starts with the definition of a given workflow, the parameters to be studied, and the input data. The workflow is then instantiated and executed efficiently in RT using parameters values selected by the SA method. These values, or parameters sets, are generated separately by the user through a SA method statically, i.e, before the execution of any task on the RTF. The output of the workflow is compared using a metric selected by the user to measure the difference between a reference segmentation result and the one computed by the workflow using the parameter set generated by the SA method. This process continues while the number of workflow runs does not achieve the sample size required by the SA. This sample size is effectively the number of times that the workflow will be instantiated and executed with different input parameters' values. The sample size is a way to limit the cost of the SA study while maintaining its significance and accuracy. This can be done by empirically choosing a sample size that is big enough to have accurate results but not enough that its cost is unfordable.

\begin{figure}[b!]
\begin{center}
\includegraphics[width=0.8\textwidth]{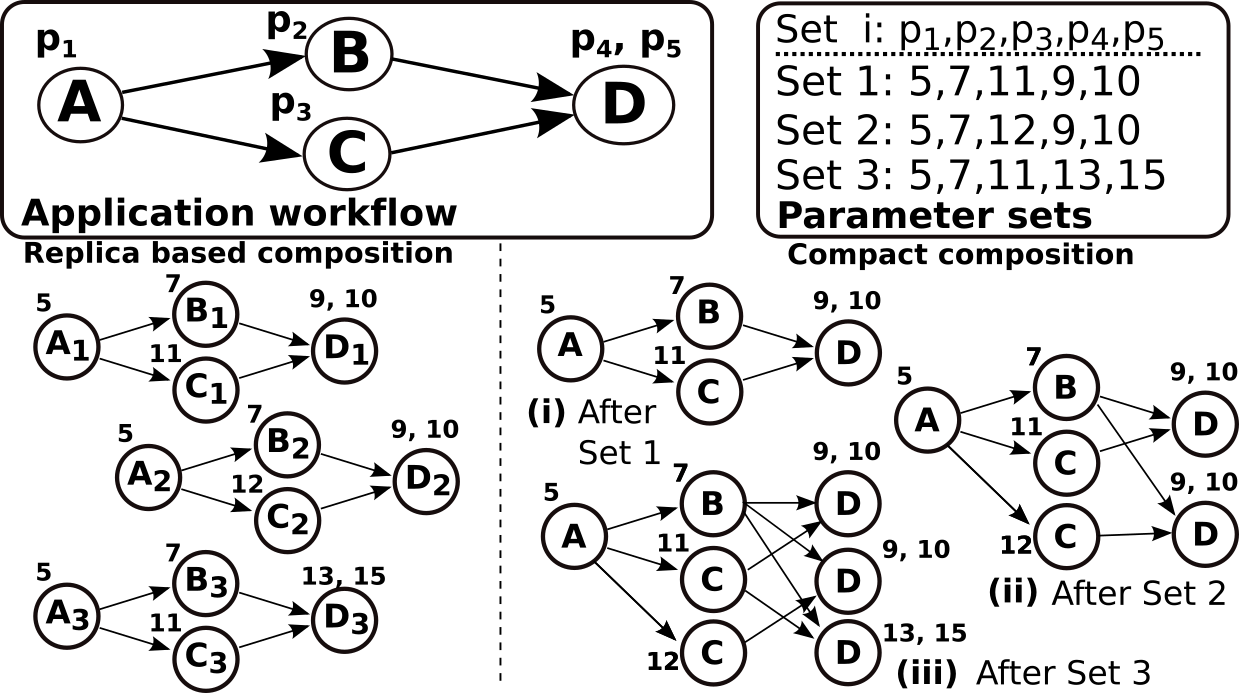}
\caption{A comparison of a workflow generated with and without computation reuse. Image extracted from \cite{rtf2}.}
\label{fig:reuse}
\end{center}
\vspace{-4mm}
\end{figure}

Computation reuse is achieved through the removal of repeated computation tasks. Figure \ref{fig:reuse} presents the comparison of a replica-based workflow generation, in which there is no reuse, and a compact composition, generated with maximal reuse. Given that we start generating a compact composition with no tasks on it, the first parameter set (Set 1, (i) in Figure \ref{fig:reuse}) is added to the workflow in its entirety (i.e., all computation tasks A-D). The second parameter set, however, has the reuse opportunities of tasks A and B given they have the same input parameters values and input data. Thus, only the tasks C and D for parameter set 2 are instantiated in the compact graph ((ii) in Figure \ref{fig:reuse}). With the current workflow state of (ii), parameter set 3 presents reuse opportunities for tasks A, B and C, thus only needing to instantiate a single computation task (D) with the parameters values 13 and 15 to the workflow. When comparing the workflow replica based composition with the compact composition we can notice a decrease on the number of executed tasks of approximately 41\%, from 12 tasks to 7 tasks.

There are two computation reuse levels used on this work, (i) stage-level, on which coarse-grain computation tasks are reused, and (ii) task-level - with fine-grain tasks reused. Coarse-grain computation reuse is significantly easier to implement than its fine-grained counterpart. However, the number of parameters that two coarse-grained merging candidates stages need to match for the reuse to take place is higher as when compared with fine-grain tasks.

\subsection{Graphical User Interface and Code Generator}
\label{sec:improve}

In this work a flexible task-based stage code generator was implemented to ease the process of developing RTF applications. This generator was created, together with a workflow generator graphical interface - with the purpose of making the RTF more accessible to domain-specific experts. Additionally, this code generator will simplify the application information gathering process, necessary for merging stages instances during the process of computation reuse.

\begin{figure}[h!]
\vspace{-4mm}
\begin{center}
\includegraphics[width=0.6\textwidth]{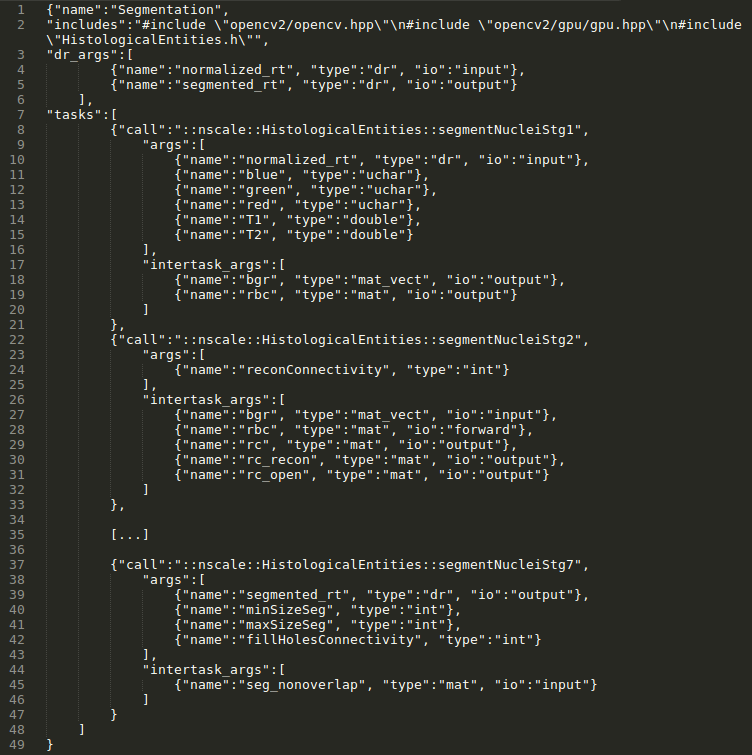}
\caption{An example stage descriptor Json file.}
\label{fig:cgen}
\end{center}
\vspace{-4mm}
\end{figure}

The stage generator has as its input a stage descriptor file, formatted as Json, as shown in Figure \ref{fig:cgen}. A stage is defined by its name, the external libraries it needs to call in order to execute the application domain transformations in each stage of the workflows, the necessary input arguments for its execution and the tasks it must execute. There are two kinds of inputs: the arguments and the Region Templates (RT). The arguments are constant inputs, which are varied by the given SA method and represent the application input parameter values. The RT is the data structure provided by the RTF for inter-stage and inter-task communication. As seen on the example descriptor file, only the RT inputs are explicitly written, while the remaining arguments can be inferred from the tasks descriptions.

Every stage is comprised of tasks, described by (i) the external call to the library of operations implemented by the user and (ii) its arguments. On Figure \ref{fig:cgen} the call for the first task is \textit{segmentNucleiStg1} from the external library \textit{nscale}. The arguments can be one of two types, (i) constant input arguments (args), defined by the SA application or (ii) intertask arguments (intertask\_args), which are produced/consumed for/by a fine-grain task.

\begin{figure}[h!]
\begin{center}
\includegraphics[width=0.8\textwidth]{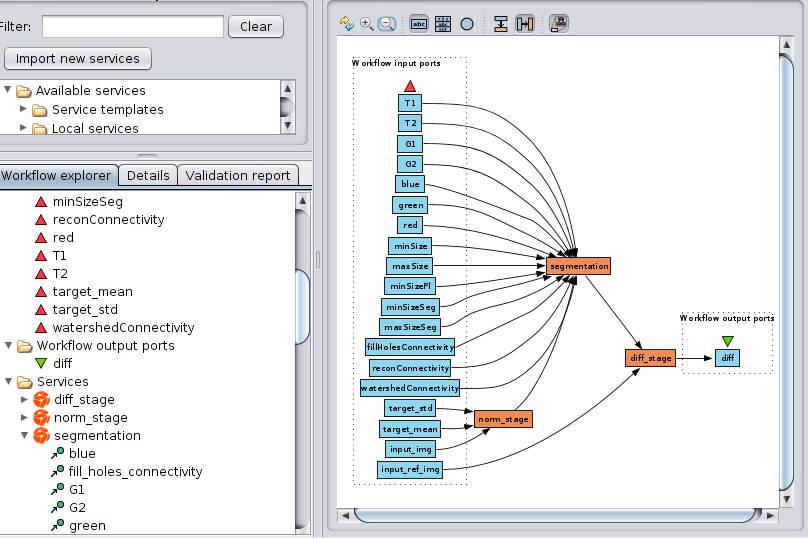}
\caption{The example workflow described with the Taverna Workbench.}
\label{fig:tav1}
\end{center}
\vspace{-4mm}
\end{figure}

With task-based stages generated, the user can instantiate workflows using the newly generated stages. As with tasks, the RTF did not support a flexible, non-compiled solution for generating workflows, being these workflows hardcoded into the RTF. The solution implemented on this work was to use the Taverna Workbench tool \cite{taverna} as a graphical interface for producing workflows and implement a parser for the generated Taverna file. An example workflow on the Taverna Workbench is displayed on Figure \ref{fig:tav1}.

\subsection{Stage-Level Merging}
\label{sec:stage-merging}

The stage level merging needs to identify and remove common stage instances and build a compact representation of the workflow, as presented in Algorithm~\ref{alg:stg-merg}. The algorithm receives the application directed workflow graph (appGraph) and parameter sets to be tested as input (parSets) and outputs the compact graph (comGraph). It iterates over each parameter set (lines 3-5) to instantiate a replica of the application workflow graph with parameters from $set$. It then calls {\scshape MergeGraph} to merge the replica to the compact representation.  

\begin{algorithm}[b!]
\small
	\caption{Compact Graph Construction
	\label{alg:stg-merg}}
		
	\begin{algorithmic}[1]
	\State {\bf Input:} appGraph; parSets;
	\State{\bf Output:} comGraph;
	\For{{\bf each} set $\in$ parSets}
		\State appGraphInst = \Call{instantiateAppGraph}{set};
		\State \Call{MergeGraph}{appGraphInst.root, comGraph.root};
	\EndFor
	\Procedure{MergeGraph}{appVer, comVer}
		\For{{\bf each} v $\in$ appVer.children}
			\If{(v' $\gets$ find(v, comVer.children))}
				\State \Call{MergeGraph}{v, v'};
			\Else
				\If{((v' $\gets$ PendingVer.find(v))==$\emptyset$)}
					\State v' $\gets$ clone(v)
					\State v'.depsSolved $\gets$ 1
					\State comVer.children.add(v')
					\If{v'.deps $\ge$ 1}
						\State PendingVer.insert(v')
					\EndIf
					\State \Call{MergeGraph}{v, v'};
				\Else
					\State comVer.children.add(v')
					\State v'.depsSolved $\gets$ v'.depsSolved+1
					\If{v'.depsSolved == v'.deps}
						\State PendingVer.remove(v')
					\EndIf
					\State \Call{MergeGraph}{v, v'}
				\EndIf
			\EndIf    
		\EndFor
	\EndProcedure
	\end{algorithmic}
\end{algorithm}

The {\scshape MergeGraph} procedure walks simultaneously in an application workflow graph instance and in the compact representation. If a path in the application workflow graph instance is not found in the latter, it is added to the compact graph.  The {\scshape MergeGraph} procedure receives the current set of vertices in the application workflow ($appVer$) and in the compact graph ($comVer$) as a parameter and, for each child vertex of the $appVer$, finds a corresponding vertex in the children of $comVer$. Each vertex in the graph has a property called {\em deps}, which refers to its number of dependencies. The find step considers the name of a stage and the parameters used by the stage. If a vertex is found, the path already exists, and the same procedure is called recursively to merge sub-graphs starting with the matched vertices (lines 9-10). When a corresponding vertex is not found in the compact graph, there are two cases to be considered (lines 11-26). In the first one, the searched node does not exist in $comGraph$. The node is created and added to the compact graph (lines 12-18). To check if this is the case, the algorithm verifies if the node ($v$) has not been already created and added to $comGraph$ as a result of processing another path of the application workflow that leads to $v$. This occurs for nodes with multiple dependencies, e.g., D in Figure~\ref{fig:reuse}. If the path (A,B,D) is first merged to the compact graph, when C is processed, it should not create another instance of D. Instead, the existing one should be added to the children list as the algorithm does in the second case (lines 21-25). The $PendingVer$ data structure is used as a look-up table to store such nodes with multiple dependencies during graph merging. This algorithm makes $k$ calls to {\scshape MergeGraph} for each $appGraphInst$ to be merged, where $k$ is the number of stages of the workflow. The cost of each call is dominated by the $find$ operation in the $comVer.children$. The $children$ will have a size of up to $n$ or $|parSets|$ in the worst case. By using a hash table to implement children, the find is $\mathcal{O}$($1$). Thus, the insertion of $n$ instances of the workflow in the compact graph is $\mathcal{O}$($kn$).

\subsection{Task-Level Merging}

On the previous section coarse-grain reuse was implemented through a stage-level merging algorithm. This approach can by itself attain good speedups for the workflow used on this work. However, due to the granularity of the stages there is still many reuse opportunities which are wasted since they are not visible or even achievable on stage-level. These opportunities are visible though on task-level, through what we define as fine-grain reuse. This reuse can be achieved by merging stages together and removing the repeated tasks, through what we call task-level merging. Merging at task-level, unlike stage-level, has some limitations due to the way stages and tasks are implemented on the RTF. Tasks are a finer-grain computational job, intended to be small activities. Although stages can be executed on distinct computing nodes, tasks cannot, since it would not make sense to distribute such small tasks which communication overhead over the nodes network would most likely outweigh the task cost itself.

With these peculiarities in mind, before we implement any fine-grain merging algorithm we must first address some limitations on excessive fine-grain reuse. When excessive task-level merging is performed the joint number of parameters and variables of a merged stage, containing a large number of tasks, may not fit on the system memory. These variables are most of the times intermediate data that is passed between tasks, also including intermediate images, which are rather large for the purpose of this work. Also, it is possible for all stages to be merged in a number smaller than the number of available nodes, hence making some of the available resources idle. Both these problems can be solved by limiting the maximum number of stages that can be merged (bucket size). This limit is defined here as $MaxBucketSize$. Another way to enforce memory restriction is to limit the maximum number of tasks per group of merged stages (buckets). This limit is the $MaxBuckets$.

\subsubsection{Na\"ive Algorithm}
\label{sec:naive-merging}

In the interest of better understanding the task-level merging problem, a na\"ive algorithm was implemented to serve as a baseline for our analysis. This simplified algorithm groups $MaxBucketSize$ stages in buckets and attempts to merge all stages of each bucket among themselves. This was achieved by sequentially grouping the first $MaxBucketSize$ stages into buckets, until there are no more stages to be merged. 


Although this simple solution was quickly implemented and has a linear algorithmic complexity its reuse efficiency is, however, highly dependent on the stages ordering. For instance, if similar stages were to be generated close together a greater amount of reusable computation is more likely to exist.

\subsubsection{Smart Cut Algorithm (SCA)}
\label{sec:sca}

\begin{figure}[!h]
	 \centering
	 \begin{subfigure}[b]{0.3\textwidth}
			\centering
			 \includegraphics[width=\textwidth]{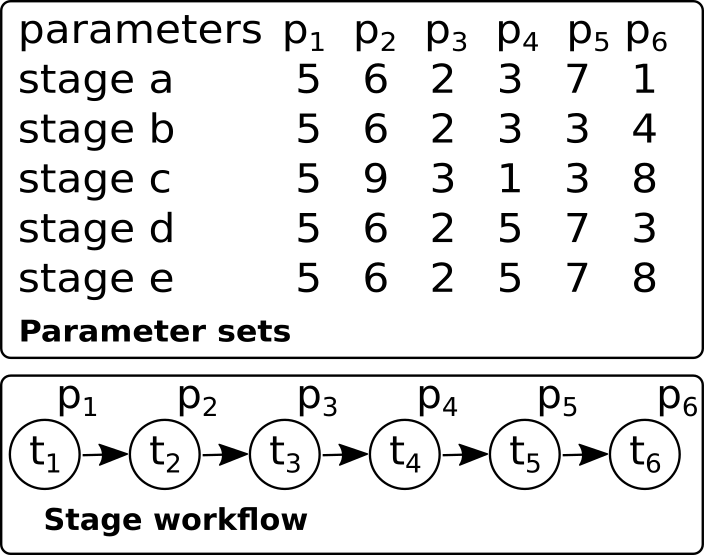}
			 \caption{Example application.}
			 \label{fig:sca0}
	 \end{subfigure}
	 \hspace{3mm}
	 \begin{subfigure}[b]{0.3\textwidth}
			\centering
			 \includegraphics[width=\textwidth]{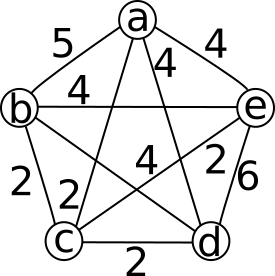}
			 \caption{Initial graph of instance example.}
			 \label{fig:sca1}
	 \end{subfigure}
	 \hspace{3mm}
	 \begin{subfigure}[b]{0.3\textwidth}
			 \centering
			 \includegraphics[width=0.8\textwidth]{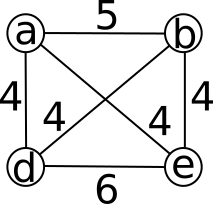}
			 \caption{First cut is performed, removing node $c$.}
			 \label{fig:sca2}
	 \end{subfigure}
	 \par\bigskip
	 \begin{subfigure}[b]{0.3\textwidth}
			 \centering
			 \includegraphics[width=0.8\textwidth]{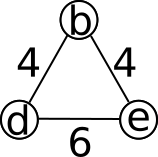}
			 \caption{After the next cut node $a$ is removed.}
			 \label{fig:sca3}
	 \end{subfigure}
	 \hspace{3mm}
	 \begin{subfigure}[b]{0.3\textwidth}
			 \includegraphics[width=0.8\textwidth]{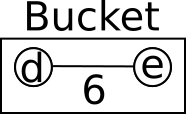}
			 \centering
			 \caption{After final cut of node $b$ $MaxBucketSize$ sized subgraph is found.}
			 \label{fig:sca4}
	 \end{subfigure}
	 \hspace{3mm}
	 \begin{subfigure}[b]{0.25\textwidth}
			 \includegraphics[width=\textwidth]{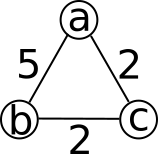}
			 \caption{The cutting starts over with the remaining nodes.}
			 \label{fig:sca5}
	 \end{subfigure}
	 
		\caption{An example on which SCA executes on 5 instances of a workflow application of 6 tasks, with $MaxBucketSize = 2$.}
		\label{fig:sca}
\end{figure}

Another strategy to create buckets of stages to be merged that was investigated is through the use of a graph based representation (see Figure \ref{fig:sca}). A representation for this could be done using fully-connected undirected graphs on which the stage instances are the nodes and each edge is the degree of reuse between two stage instances (Figure \ref{fig:sca1}). By degree of reuse we mean the number of tasks that would be reused if the two stages are merged. With this perspective we would need only to partition this graph in subgraphs, maximizing the reuse degree of all subgraphs. This is a well-known problem, called min-cut \cite{mincut}. 

Although there are many variations for the min-cut problem \cite{mincut,minkcut}, we define here a min-cut algorithm as one that takes an undirected graph and performs a 2-cut (i.e., cut the graph in two subgraphs) operation, minimizing the sum of the cut edges weight. This 2-cut operation was selected because of its flexibility and computational complexity. First, the recursive use of 2-cuts can break a graph in any number of subgraphs. Moreover, k-cut algorithms are not only more computationally intensive than 2-cut algorithms, but also have no guarantees for the balancing of the subgraphs (e.g., for $k=5$ on a graph with 10 nodes one possible solution is 4 subgraphs with 1 node each and 1 subgraph with 6 nodes). As such, we can implement a simple k-cut balanced algorithm by performing 2-cut operations on the most expensive graph/subgraph until a stopping condition is reached (e.g., number of subgraphs is reached, number of nodes per subgraph is reached). With all these considerations only 2-cut operations are used on the proposed algorithm.

Figure \ref{fig:sca} demonstrates a way to group stages into buckets using 2-cut operations. First, the fully-connected graph in Figure \ref{fig:sca1} is generated given the stage instances of Figure \ref{fig:sca0}. Figure \ref{fig:sca2} shows the result of the first 2-cut operation, on which the subgraph containing only the node $c$ is found to be the one least related to the subgraph with the remaining nodes. This is similar to the state that $c$ is the ``least reusable'' stage among all other stages (i.e., the stage which, if selected for merging, would have highest computational cost). Next, nodes $a$ and $b$ are removed until a bucket of size 2 is reached (see Figures \ref{fig:sca2} and \ref{fig:sca3}). The previously removed nodes ($a$, $b$ and $c$) are then put together (Figure \ref{fig:sca5}) and the same cutting algorithm starts over. This process is then repeated until all stages are grouped into buckets.

With this procedure in mind Algorithm \ref{alg:scut} was designed. 
This algorithm performs successive 2-cut operations on the graphs to divide it into
disconnected subgraphs that fit in a bucket. The cuts are performed such that
the amount of reuse lost with a cut is minimized. In more detail, the partition
process starts by dividing the graph into 2 subgraphs (s1 and s2) using a minimum cut
algorithm~\cite{mincut} (line 4). Still, after the cut, both
subgraphs may have more than $MaxBucketSize$ vertices. In this case, another
cut is applied in the subgraph with the largest number of stages (lines 5-7), and this is
repeated until a viable subgraph (number of stages $\leq$ $MaxBucketSize$) is
found.  When this occurs, the viable subgraph is removed from the original
graph (lines 8-11), and the full process is repeated until the graphs with stage instances
yet not assigned to a bucket can fit in one.

\begin{algorithm}[t!]
\small
	\caption{Smart Cut Algorithm
		\label{alg:scut}}
				
	\begin{algorithmic}[1]
		\State {\bf Input:} stages; MaxBucketSize;
	\State{\bf Output:} bucketList;
		
		\While{|stages| > 0}
			\State \{s1,s2\} $\gets$ \Call{2cut}{stages}
				\While{|s1| > MaxBucketSize}
					\State \{s1,s2\} $\gets$ \Call{2cut}{s1}
				\EndWhile
				\State bucketList.add(s1)
				\For{{\bf each} s $\in$ s1}
					\State stages.remove(s)
				\EndFor
		\EndWhile
	\end{algorithmic}
\end{algorithm}

The number of cuts necessary to compute a single viable subgraph of $n$ stages is $\mathcal{O}$($n$) in the worst case. This occurs when each cut returns a subgraph with only one stage and another subgraph with the remaining nodes. The cut then needs to be recomputed -- about $n-MaxBucketSize)$ times -- on the largest subgraph until a viable subgraph is found. Also, in the worst case, all viable subgraphs would have $MaxBucketSize$ stages and, as such, up to $n/MaxBucketSize$ buckets could be created. Therefore, the algorithm will perform $\mathcal{O}$($n^2$) cuts in the worst case to create all buckets. In our implementation, the min-cut is computed using a Fibonacci heap~\cite{mincut} to speed up the algorithm, making each cut $\mathcal{O}$($E+V\ log\ V$). Since the graph used is fully connected, the complexity of a single cut in our case is $\mathcal{O}$($n^2$) and, as consequence, the full SCA is $\mathcal{O}$($n^4$).  Although the SCA computes good reuse solutions, its use in practice is limited because of the computational complexity.  This motivated the proposal of the strategy described in Section \ref{sec:rtma}.




\subsubsection{Reuse Tree Merging Algorithm (RTMA)}
\label{sec:rtma}

Still on graphs, a natural way to display hierarchical structures is with trees. Using tasks as nodes on this tree, subtrees with the same parent node indicates that all child task nodes of said parent node use its output. As such, if we constructed a tree with several stages, we are able to easily see the reuse opportunities, lying in the nodes with more than one child node. Moreover, each level of the tree would represent a given task, which can be instantiated with different parameters sets. 

Detailing this structure, each level of the tree represents a task, and if a stage $s$ shares a parent node on level $k$ with $s'$, this implies that all tasks from 1 to $k$ are the same for both stages, an thus reusable among themselves (i.e., same computational task with the same inputs). This structure is defined as a Reuse Tree, with every node being defined by its level (or height), its parent, its children and a reference to the stage responsible for its generation.

\paragraph{Reuse-Tree Generation}

On a SA example we have a workflow $w$ that is instantiated $n$ times with different parameters ($w_1$, $w_2$, ..., $w_n$). Each workflow $w_i$ is composed by $m$ stages $s_{ij}$ with $i \in [1,n]$ and $j \in [1,m]$. A reuse-tree is then generated for each $j$-th stage level. The reuse-tree for a given stage level can be generated by iteratively inserting one stage instance after the other on the reuse-tree. Initially, a stage is represented as a tree on which every node has a single child and each node represents a task instantiation for that stage. Furthermore, any given node has as its parent a task that it is dependent on. Every stage is inserted one task node at a time. If, for a given task node, there already exists on the tree another node representing the same task with the same parameter inputs, said task node is not created, but instead the insertion process carries on from the equivalent node, characterizing task reuse.

\begin{figure}[h]
	 \centering
	 \begin{subfigure}[b]{0.3\textwidth}
			\centering
			 \includegraphics[width=\textwidth]{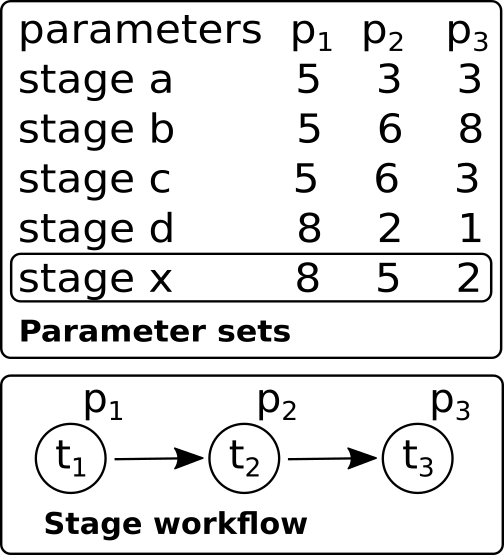}
			 \caption{Example application.}
			 \label{fig:gen1}
	 \end{subfigure}
	 \hspace{3mm}
	 \begin{subfigure}[b]{0.3\textwidth}
			 \centering
			 \includegraphics[width=0.8\textwidth]{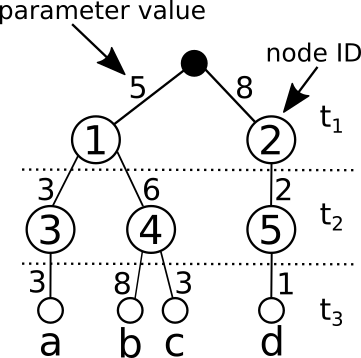}
			 \caption{Initial reuse tree for the instance example.}
			 \label{fig:gen2}
	 \end{subfigure}
	 \hspace{3mm}
	 \begin{subfigure}[b]{0.3\textwidth}
			 \centering
			 \includegraphics[width=0.8\textwidth]{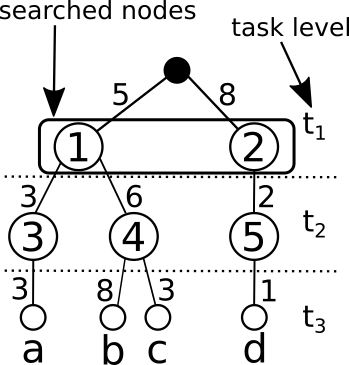}
			 \caption{Searching for reuse on the first task.}
			 \label{fig:gen3}
	 \end{subfigure}
	 \par\bigskip
	 \begin{subfigure}[b]{0.3\textwidth}
			 \includegraphics[width=0.8\textwidth]{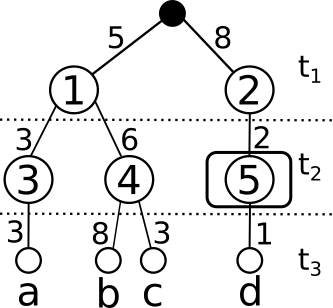}
			 \centering
			 \caption{Searching for reuse on the second task.}
			 \label{fig:gen4}
	 \end{subfigure}
	 \hspace{3mm}
	 \begin{subfigure}[b]{0.3\textwidth}
			 \includegraphics[width=\textwidth]{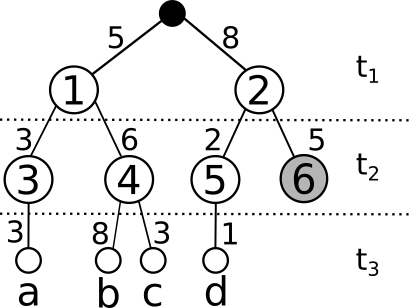}
			 \caption{Inserting a new node, 6.}
			 \label{fig:gen5}
	 \end{subfigure}
	 \hspace{3mm}
	 \begin{subfigure}[b]{0.3\textwidth}
			 \includegraphics[width=\textwidth]{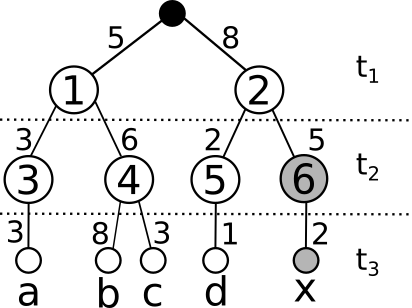}
			 \caption{Inserting the leaf node $x$.}
			 \label{fig:gen6}
	 \end{subfigure}
	 
		\caption{An example where node $x$ is inserted on the existing reuse tree. Figure \ref{fig:gen1} defines the tasks of which each stage is composed by and presents the parameters' values for each stage instance.}
		\label{fig:gen-ex}
\end{figure}

As an example, Figure \ref{fig:gen-ex} demonstrates the insertion of a stage (stage $x$) with the stage workflow and the parameters of each stage instance defined in Figure \ref{fig:gen1}, and the starting reuse tree in Figure \ref{fig:gen2}. Starting at the root node, its children (1 and 2) are searched for reuse opportunities for the first task (Figure \ref{fig:gen3}). Since node 2 represents all stages whose task 1 has as its input $p_1 = 8$ the first task of $x$ can be reused through it. The search for reuse of the second task is then performed on the children of node 2 (Figure \ref{fig:gen4}). Since node's 2 only child, node 5, cannot be reused for stage $x$'s second task (values for $p_2$ of stages $d$ and $x$ are different), a new node representing this non-reusable task is created (node 6) as shown in Figure \ref{fig:gen5}. Finally, since node 6 is new, there are no more reuse opportunities from it, thereby, a single child node must be created for each of the remaining non-reusable tasks (Figure \ref{fig:gen6}).

\paragraph{The Merging Implementation}

In order for a merging algorithm to be implemented on top of the Reuse Tree structure we must take advantage of its hierarchical characteristics. Given that we want to bundle together buckets of stages of exactly $MaxBucketSize$ stages we must start with the deepest stages and move up. Figure \ref{fig:rg1} shows an example of a Reuse Tree with 12 stages and 3 tasks each. Stages $a$, $b$ and $c$ have two out of three reusable tasks, and as such, given a $MaxBucketSize=3$, should be put together in the same bucket. Meanwhile, stages $d$ through $i$ have one out of three reusable tasks. To maximize the reuse, stages $d$, $e$, $f$ and $g$ should be together, as should stages $h$ and $i$. Since $MaxBucketSize=3$, only 3 stages out of $d$, $e$, $f$ and $g$ can be put together, not mattering which 3 stages. This merger is seen in Figure \ref{fig:rg2}. After the merger of $d$, $e$ and $f$, stage $g$ is left alone, having as its best option, reuse-wise, to be put together with $h$ and $i$. As such, it is visible that the merging should happen in a bottom-up fashion.

\begin{figure}[!h]
	 \centering
	 \begin{subfigure}[b]{0.45\textwidth}
			 \includegraphics[width=\textwidth]{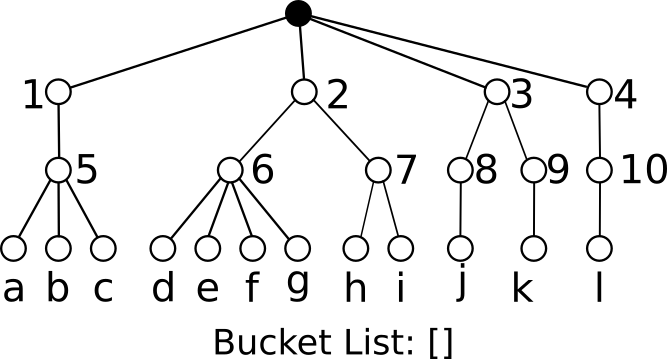}
			 \caption{Initial reuse tree.}
			 \label{fig:rg1}
	 \end{subfigure}
	 \hspace{3mm}
	 \begin{subfigure}[b]{0.45\textwidth}
				\includegraphics[width=\textwidth]{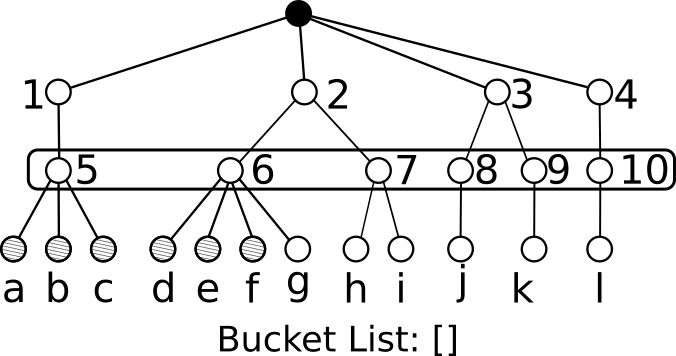}
				\caption{Reuse Tree after select procedure.}
				\label{fig:rg2}
		\end{subfigure}
		\par\bigskip
			 \begin{subfigure}[b]{0.45\textwidth}
				\includegraphics[width=\textwidth]{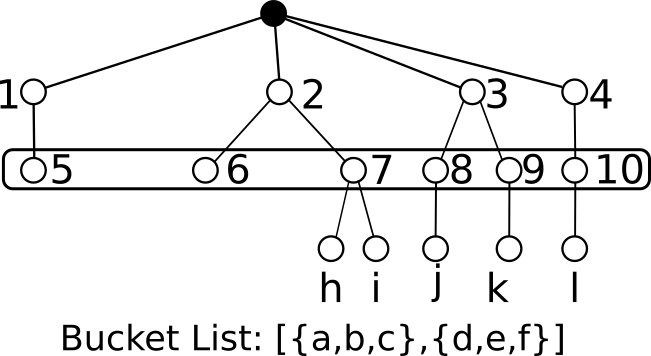}
				\caption{Reuse Tree after the selected mergeable leaf nodes are pruned and added to the bucket list.}
				\label{fig:rg2}
		\end{subfigure}
		\hspace{3mm}
		\begin{subfigure}[b]{0.45\textwidth}
				\includegraphics[width=\textwidth]{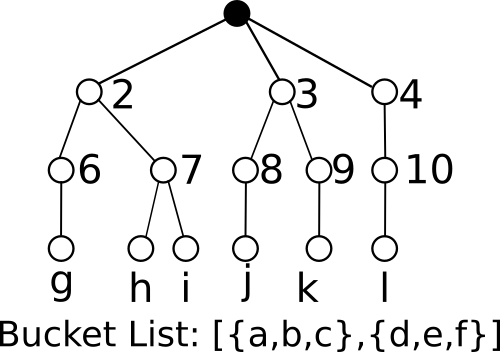}
				\caption{Reuse tree after the childless parents are recursively removed.}
				\label{fig:rg3}
		\end{subfigure}
		\par\bigskip
		\begin{subfigure}[b]{0.45\textwidth}
				\includegraphics[width=\textwidth]{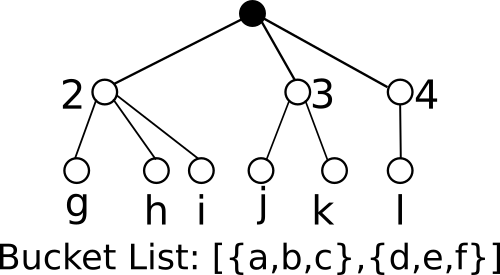}
				\caption{Reuse tree after move-up procedure.}
				\label{fig:rg4}
		\end{subfigure}
		\caption{An example of Reuse Tree based merging with MaxBucketSize 3. The merged stages of each step are shown below the tree on the bucket list.}
		\label{fig:rt-example}
\end{figure}

\begin{algorithm}
\footnotesize
	\caption{Reuse-Tree Merging Algorithm (RTMA)\label{alg:rtm}}

	\begin{algorithmic}[1]
	\State {\bf Input:} stages; maxBucketSize;
	\State{\bf Output:} bucketList;

	\State bucketList $\gets \emptyset$; 

	\State rTree $\gets$ \Call{GenerateReuseTree}{stages}

	\While{rTree.height $>$ 2}
		\State leafsPList $\gets$ \Call{GenerateLeafsParentList}{rTree}
		\State newBuckets $\gets$ \Call{PruneLeafLevel}{rTree, leafsPList, maxBucketSize}
		\State bucketList $\gets$ bucketList $\cup$ newBuckets
		\State \Call{MoveReuseTreeUp}{reuseTree, leafsPList}
	\EndWhile
	\While{rTree.root.children $\neq \emptyset$}
		\State newBucket $\gets \emptyset$
		\State newBucket.add(removeFirstChildren(rTree.root.children));
		\State bucketList $\gets$ bucketList $\cup$ newBucket
	\EndWhile
	\State \Return bucketList

	\end{algorithmic}
\end{algorithm}

The Reuse Tree Merging Algorithm (RTMA), listed on Algorithm \ref{alg:rtm}, was implemented in three steps, (i) bucket candidates selection (line 6), (ii) tree pruning (line 7) and (iii) move-up operation (line 9), which are performed iteratively until the whole tree is consumed. If at the end of the main loop (line 5) there are still any non mergeable stages, those will be converted to one-stage buckets (lines 12-13) and then inserted on the final solution (line 14).

The first step of the algorithm (Algorithm \ref{alg:rtm}, line 6) is to get a list of all parents of leaf nodes. In Figure \ref{fig:rg2} we can see the selected parents (5-10). With the leaf's parents list the pruning step makes as many $MaxBucketSize$ sized buckets as possible and then remove them from the reuse tree. The procedure $PruneLeafLevel$ (line 7) attempts to make buckets for each leaf parent node. As stated before, the new buckets must have an exact size of $MaxBucketSize$, thereby, if the parent node does not have at least $MaxBucketSize$ children will not create a bucket with them. Given that the parent has enough children, a number of $MaxBucketSize$ children will be bundled together as a new bucket to later be added to the solution pool. On Figure \ref{fig:rg2} the two formed buckets are shown: $\{a,b,c\}$ and $\{d,e,f\}$. Each time a leaf node is added to the current new bucket, it is then removed from the parent children list, and as a consequence, removed from the tree, as seen on Figure \ref{fig:rg3}.

If a parent node ends up grouping all of its children in buckets, it must be removed from the tree (node 5 on Figure \ref{fig:rg2}). This process is performed recursively by removing the given childless parent node and then checking if the removal of the current parent also makes its parent childless. If this is the case the parent node removal must continue on its parent (node 1 of Figure \ref{fig:rg2} is also removed, as seen on Figure \ref{fig:rg3}).






The final step of merging is to move the leaf nodes up one level in order to enable the creation of new buckets. The operation $MoveReuseTreeUp$ (Algorithm \ref{alg:rtm}, line 9) is done by taking each of the previously selected parent nodes and moving all of its children to its parent's children list (e.g., nodes $g$, $h$ and $i$ of Figure \ref{fig:rg3} are moved to parent node 2, as seen on Figure \ref{fig:rg4}). After that, the current node is remove from its parent (e.g., nodes 6 and 7 of Figure \ref{fig:rg3} are removed from parent node 2, as seen on Figure \ref{fig:rg4}). After all nodes from the parent list are removed and its children are moved up the tree height is updated (line 6).

\paragraph{Algorithmic Complexity}

Assuming an empty tree, the {\scshape GenerateReuseTree} performs the insertion of $n$ stage instances with $k$ tasks each. In the worst case of a stage insertion there is no reuse whatsoever, resulting in the creation of the maximum number of nodes. In this case, given that $m<n$ stage instances were already added, the next stage will perform $m$ comparisons, looking for a reuse opportunity. After no opportunities are found $k$ nodes will be created. This results in $kn$ new nodes generated and $n(n+1)/2$ nodes traversed in total, and as a consequence, {\scshape GenerateReuseTree} is $\mathcal{O}$($n^2$).

The $n(n+1)/2$ nodes traversed is due to a linear search for reusable tasks on a given level with $m$ stages instances. It is possible to further reduce this cost by performing this reuse check on a hash table on which the key is a combination of all parameters' values. By doing this hash table search the cost of each insertion will be $\mathcal{O}$(1), thus resulting in the overall time complexity of $\mathcal{O}$($kn$).


The analysis of the actual merging algorithm can be split in the three operations performed on the reuse tree. Starting with the select operation, on the worst case, there will be one child per stage (i.e., no reuse on the first task), resulting in $n$ nodes visited. On this case, the number of children of each node beyond the first level will be one, resulting in $k-2$ extra nodes visited. As a result we have that that $GenerateLeafsParentList$ runs in $\mathcal{O}$($nk$) per iteration, or $\mathcal{O}$($nk^2$), for there are exactly $k-1$ iterations.

For the pruning step the most expensive operation is the one of adding a stage to a solution bucket. Knowing that the exact number of bucket insertions must be at most $n$ for the whole merging algorithm, we get the complexity $O(n)$ for all iterations of the pruning step. 

At last, the complexity of the move-up step will be calculated by the amount of times a leaf node is moved from the current node child list to its parent. Independently to the structure of the tree, given that it has $n$ leaf nodes, all of them will be moved once per move-up operation. Given that there are exactly $k-1$ iterations, we have $\mathcal{O}$($nk$).

The RTMA complexity is then dominated by tree generation algorithm since it is $\mathcal{O}$($n^2$), versus the joint complexity of the other three steps, $\mathcal{O}$($nk^2 + nk + n$). This happens because $n \gg k$ by the order of hundreds to thousands times greater. With such time complexity, the RTMA is expected to be scalable enough in order to be a viable solution. Furthermore, if the hash table optimization of {\scshape GenerateReuseTree} is implemented, then the time complexity becomes $\mathcal{O}$($nk^2$).

\subsubsection{Task-Balanced Reuse-Tree Merging Algorithm (TRTMA)}
\label{sec:TRTMA}

Given the nature of the chosen SA application and its scale (in terms of compute cost and resources utilization), if the scale of resources is high enough, or the chosen SA method requires a sample size small enough, the ratio of buckets per computing node (or core) may become low. This may lead to imbalance, and thereafter performance degradation. This happens since the RTMA naturally reduces the parallelism of the application due to its grouping of stages. To solve this problem, a task-wise balanced version of RTMA was designed and implemented, the Task-Balanced Reuse-Tree Merging Algorithm (TRTMA). The TRTMA will be presented in five parts. First a general idea of the imbalance problem and how to solve it is presented. Then, algorithmic details are presented, followed by the complexity analysis of the TRTMA. Finally, some optimizations are described along with a qualitative analysis of the achievable results of the TRTMA.

\paragraph{General Idea}

In more details, the RTMA balances its buckets stage-wise. This means that the buckets generated by it have similar (or most of the times, the same) number of stages. As such, buckets imbalance comes from the difference on the number of tasks that two buckets with the same number of stages can have. This difference is a consequence of distinct reuse patterns on a reuse-tree structure, which in turn leads to different numbers of tasks for buckets with the same number of stages.

Given this imbalance of stage-wise balanced buckets, the TRTMA can be seen as an improvement of the RTMA on which task-wise balancing is enforced. In order to do so, the TRTMA also uses the reuse-tree structure, while trying to achieve the best balancing for $MaxBuckets$ buckets. The change of the $MaxBucketSize$ parameter to $MaxBuckets$ helps the usability of the algorithm since the maximum number of buckets is a higher-level concept than the maximum number of stages per buckets.

The TRTMA is implemented in three steps. On the first two, the $MaxBuckets$ number of buckets is achieved to then be balanced task-wise on a third step. The tree steps are defined as: Full-Merge, Fold-Merge and Balance.

\begin{figure}[h!]
	 \centering
	 \begin{subfigure}[t]{0.22\textwidth}
			 \includegraphics[width=0.9\textwidth]{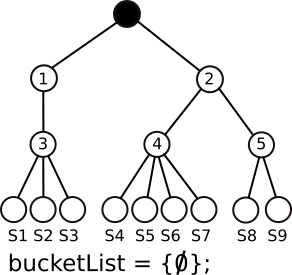}
			 \caption{Initial reuse-tree.}
			 \label{fig:trma-full-merge-p1}
	 \end{subfigure}
	 \hspace{3mm}
	 \begin{subfigure}[t]{0.22\textwidth}
				\includegraphics[width=0.9\textwidth]{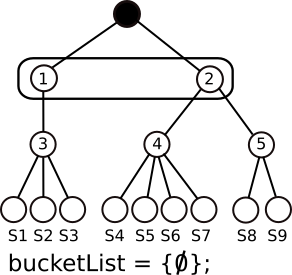}
				\caption{Attempt of {\em Full-Merge} from root node results in two buckets.}
				\label{fig:trma-full-merge-p2}
		\end{subfigure}
		\hspace{3mm}
		\begin{subfigure}[t]{0.22\textwidth}
				\includegraphics[width=0.9\textwidth]{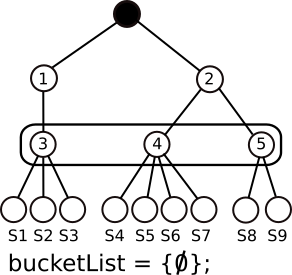}
				\caption{Attempt of {\em Full-Merge} on the children of previous attempt results in three buckets.}
				\label{fig:trma-full-merge-p3}
		\end{subfigure}
		\hspace{3mm}
		\begin{subfigure}[t]{0.22\textwidth}
				\includegraphics[width=0.9\textwidth]{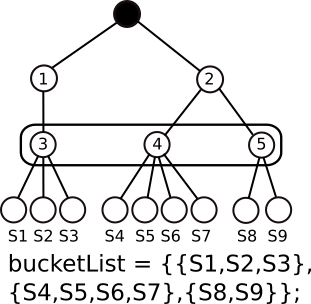}
				\caption{Merger is performed since the correct number of buckets was achieved.}
				\label{fig:trma-full-merge-p4}
		\end{subfigure}
		\caption{Simple example of {\em Full-Merge} on which $MaxBuckets$ is 3 and the exact division os stages is reached.}
		\label{fig:trma-full-merge-p}
		\vspace{-3mm}
\end{figure}

\textbf{Full Merge} is the first attempt at achieving $MaxBuckets$ buckets. It is done by traversing the reuse-tree on a top-down fashion, attempting to find a task-level on which there are at least $MaxBuckets$ nodes. The full process can be seen on Figure \ref{fig:trma-full-merge-p}, on which $MaxBuckets=3$ is used. Figure \ref{fig:trma-full-merge-p1} shows a simple initial reuse-tree. On the first level of tasks there are only two nodes (1 and 2), meaning that the next level should be searched (see Figure \ref{fig:trma-full-merge-p2}). The next level has the exact number of tasks (nodes 2, 4 and 5) and, as such, the buckets can be generated by the leaf-nodes on branches at this level (see Figure \ref{fig:trma-full-merge-p3}). Finally, the buckets are generated on Figure \ref{fig:trma-full-merge-p4}.

\begin{figure}[t!]
	 \centering
	 \begin{subfigure}[t]{0.3\textwidth}
			 \includegraphics[width=\textwidth]{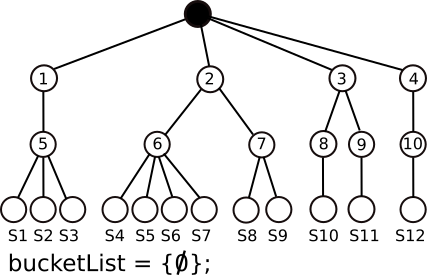}
			 \caption{Initial reuse-tree.}
			 \label{fig:trma-full-merge1}
	 \end{subfigure}
	 \hspace{1mm}
	 \begin{subfigure}[t]{0.3\textwidth}
				\includegraphics[width=\textwidth]{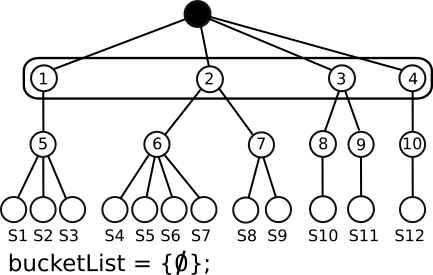}
				\caption{Attempt of {\em Full-Merge} from root node results in four buckets.}
				\label{fig:trma-full-merge2}
		\end{subfigure}
		\hspace{1mm}
		\begin{subfigure}[t]{0.3\textwidth}
				\includegraphics[width=\textwidth]{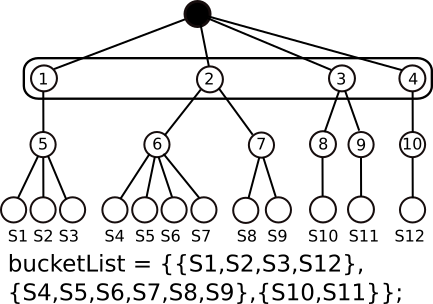}
				\caption{After {\em Fold-Merge} the correct number of buckets is reached.}
				\label{fig:trma-full-merge3}
		\end{subfigure}
		\caption{Another example of {\em Full-Merge} and {\em Fold-Merge} on which $MaxBuckets$ is 3 and the exact division of stages cannot be reached by {\em Full-Merge}.}
		\label{fig:trma-full-merge}
		\vspace{-3mm}

\end{figure}

\begin{figure*}[t!]
\begin{center}
				\includegraphics[width=0.65\textwidth]{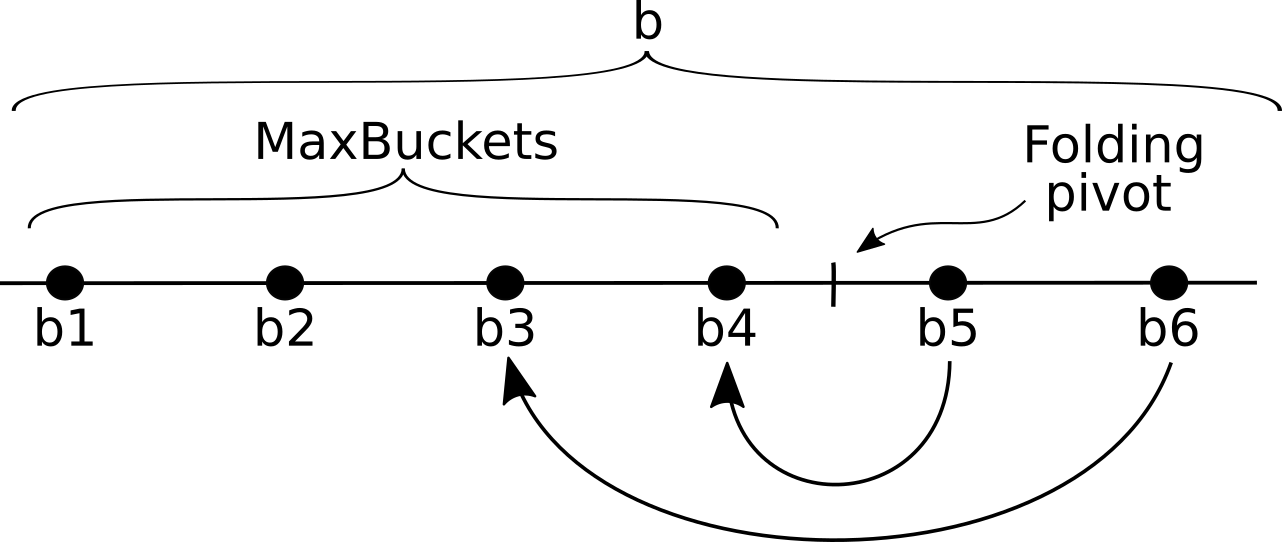}
\caption{An example of a Fold-Merging of buckets b1-b6. Initially we start with $b=6$ buckets, trying to achieve $Mb=4$ buckets. In order to do so $b-Mb$ merger operations are performed. The task cost of the buckets follows the ordering $b1 \geq b2 \geq b3 \geq b4 \geq b5 \geq b6$.}
\label{fig:fold}
\end{center}
\vspace{-4mm}
\end{figure*}

However, there are cases on which a perfect number of $MaxBuckets$ cannot be achieved (see Figure \ref{fig:trma-full-merge}). On these cases, the Full-Merge step brakes stages in a number of buckets greater than $MaxBuckets$ (see Figure \ref{fig:trma-full-merge2}). The $MaxBuckets$ number of buckets is then achieved by the merging of $b-Mb$ buckets, with $b$ being the current number of buckets and $Mb$ the $MaxBuckets$ goal on the next step (see Figure \ref{fig:fold}). 

\textbf{Fold-Merge}, as demonstrated on Figure \ref{fig:fold}, merges the buckets with the smallest cost in a fold-like operation. Given that the buckets are sorted by decreasing order, according to the cost (number of tasks), we can imagine that a line of these buckets is folded on a \textit{folding pivot}, between $Mb$ and $Mb+1$ (see Figure \ref{fig:fold}), with $Mb$ being the $MaxBuckets$ value. By doing this we are reducing the maximum bucket cost of the merged buckets, and thus reducing the amount of imbalance. It is important to notice that although the folding-fashion on which the buckets are merged is not necessary, its use mitigates the initial imbalance of the $MaxBuckets$ buckets, therefore reducing the cost of balancing these buckets. 

On the example of Figure \ref{fig:trma-full-merge2} four buckets are achieved through the Full-Merge procedure. As such, the Fold-Merge would then take the two last buckets and merge them together, resulting in the buckets of Figure \ref{fig:trma-full-merge3}.

\textbf{Balance} is the last step of the TRTMA. The balancing of buckets is done by searching for improvement operations on the reuse-tree grouped in the initial buckets outputted by the previous steps. An improvement operation is defined as a node of the reuse-tree ($imp$), which leaf nodes (or stages) can be sent from an original big reuse-tree node ($bigRT$) to a small one ($smallRT$), resulting in $newImbal \leq oldImbal$, with $newImbal$ being the imbalance on the number of tasks after the $imp$ operation, and $oldImbal$ the initial imbalance before the $imp$ operation.


\begin{figure}[t!]
	 \centering
	 \begin{subfigure}[t]{0.3\textwidth}
			 \centering
			 \includegraphics[width=\textwidth]{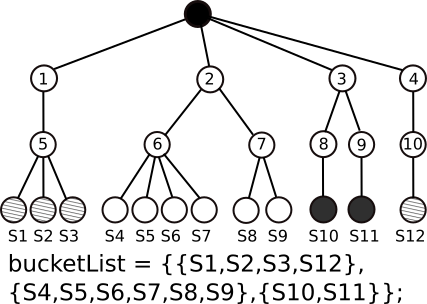}
			 \caption{Initial reuse-tree with 3 buckets of costs 8, 9 and 5, respectively.}
			 \label{fig:balance1}
	 \end{subfigure}
	 \hspace{1mm}
	 \begin{subfigure}[t]{0.3\textwidth}
			 \centering
			 \includegraphics[width=\textwidth]{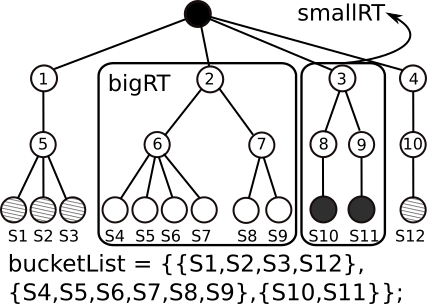}
			 \caption{Buckets of greatest and smallest cost values are selected for balance, with current imbalance of 4 and max cost 9.}
			 \label{fig:balance2}
	 \end{subfigure}
	 \hspace{1mm}
	 \begin{subfigure}[t]{0.3\textwidth}
			 \centering
			 \includegraphics[width=\textwidth]{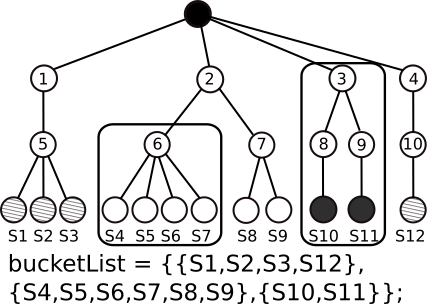}
			 \caption{The balance operation of sending node 6 children to smallRT is attempted, resulting in an imbalance of 7.}
			 \label{fig:balance3}
	 \end{subfigure}
	 \par\bigskip
	 \begin{subfigure}[t]{0.3\textwidth}
			 \centering
			 \includegraphics[width=\textwidth]{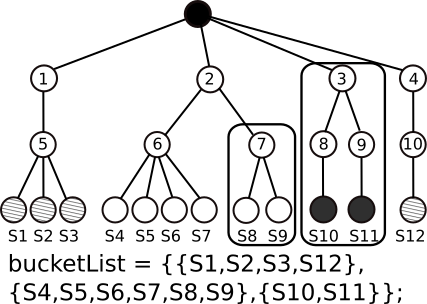}
			 \caption{After the bad selection of node 6, balancing with node 7 is also done, resulting in an imbalance of 3 but still having a max cost 9.}
			 \label{fig:balance4}
	 \end{subfigure}
	 \hspace{1mm}
	 \begin{subfigure}[t]{0.3\textwidth}
			 \centering
			 \includegraphics[width=\textwidth]{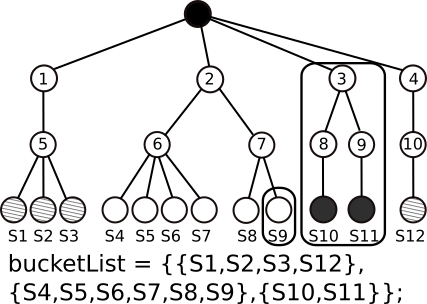}
			 \caption{As a final attempt, by sending node S9 to smallRT we have an imbalance of 0 and max cost 8, making it a viable balancing operation.}
			 \label{fig:balance5}
	 \end{subfigure}
	 \hspace{1mm}
	 \begin{subfigure}[t]{0.3\textwidth}
			 \centering
			 \includegraphics[width=\textwidth]{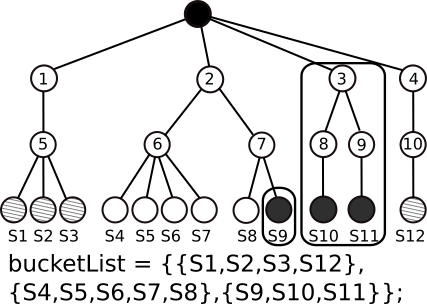}
			 \caption{After the balancing operation of sending node S9 to smallRT we have the buckets with updated costs 8, 8 and 8.}
			 \label{fig:balance6}
	 \end{subfigure}
	 \caption{An example of the {\em Balance} step on which there are 3 buckets to be balanced.}
	 \label{fig:balance}
\end{figure}

However, there are cases on which an improvement is found but it cannot positively impact the overall solution of buckets. For example, the cost $TaskCost(smallRT \cup imp)$ may be the same as $TaskCost(bigRT)$, meaning that $imp$ had some degree of reuse with $bigRT$ and thus, the cost of $imp$ is greater on $smallRT$ than on $bigRT$. On this case this improvement will reduce the imbalance but will not reduce the makespan of the application (i.e., the maximum number of tasks of all buckets). This is defined as a false improvement, or false balance operation, and since these can only increase the overall application cost, they will never be applied. As an example, two buckets $b1$ and $b2$ have initial costs 4 and 7, and thus 3 tasks of imbalance. After a given improvement their costs are 7 and 5 respectively. The imbalance is now of 2 tasks, which may be ``better'', but does not improves the makespan, which remains the same.

The full balance of a pair of buckets is achieved by attempting to find and applying valid improvements until there are no more improvements. Each improvement-search iteration is executed for a pair of the two ends of the current $bucketList$, which should be sorted in a non-ascending order with respect to cost (number of unique tasks). These are defined as $bigRT$ and $smallRT$, and are the buckets with the greatest and smallest task costs respectively. If a single improvement attempt fails, then the balance step finishes. This single improvement attempt operation is defined as {\em SingleBalance}.

A {\em SingleBalance} operation consists in traversing the $bigRT$ subtree in a breadth-first, bottom-up fashion in the search for a node that can be sent from $bigRT$ to $smallRT$, characterizing an improvement. The reason for this traversal order is that lower nodes on the reuse-tree will have at most the same number of leaf nodes of its parent and thus, finer-grain nodes are balanced earlier.

The full {\em Balance} process is exemplified on Figure \ref{fig:balance}. Starting with an initial reuse-tree of Figure \ref{fig:balance1}, the $bigRT$ and $smallRT$ buckets are selected, with task costs 9 and 5, respectively, and thus resulting in an imbalance of 4 (see Figure \ref{fig:balance2}). Several nodes are searched as improvement candidates. When trying an improvement operation of node 6, the resulting buckets $\{S8,S9\}$ and $\{S4,S5,S6,S7,S10,S11\}$ would have costs 4 and 11, resulting in a new imbalance of 7, making this operation impracticable (see Figure \ref{fig:balance3}). Another alternative is the improvement of node 7. This results in buckets $\{S4,S5,S6,S7\}$ and $\{S8,S9,S10,S11\}$, and costs 6 and 9. This improvement operation results in an imbalance of 3, which is better than the previous imbalance of 4. However, the maximum task cost still remains at 9, meaning this is a false improvement and hence, an invalid balancing operation since the makespan has not changed (see Figure \ref{fig:balance4}). Finally, by applying the improvement of leaf-node $S9$ (which could be any of the nodes in the interval $[S4,S9]$) the resulting buckets would be $\{S4,S5,S6,S7,S8\}$ and $\{S9,S10,S11\}$, both with cost 8 and thus, 0 of imbalance (see Figure \ref{fig:balance5}). Given that this last improvement operation was the best found, it is applied (see Figure \ref{fig:balance6}). Since it is impossible to improve the imbalance of 0, the next {\em SingleBalance} attempt will not find any valid improvement, consequently ending the {\em Balance} step.

\paragraph{Algorithmic Implementation Details}

On this section the {\em Balance} and {\em SingleBalance} algorithms will be detailed since they are the most complex of all previously presented algorithms. Going through in a bottom-up fashion, {\em SingleBalance} is detailed in Algorithm \ref{alg:balance_single}.

\begin{algorithm}[!h]
\footnotesize
		\caption{Balancing algorithm for two tree nodes ({\em SingleBalance})\label{alg:balance_single}}

		\begin{algorithmic}[1]
		\State {\bf Input:} currChildren; bigRT; smallRT; imbal;
		\State{\bf Output:} improvement;

		\While{$|$currChildren$|$ = 1 {\bf and} $|$ currChildren.first().children$|$ > 0}
			\State currChildren $\gets$ currChildren.first()
		\EndWhile

		\State uniqueChildren $\gets$ $\emptyset$
		\State uniqueChildrenCosts $\gets$ $\emptyset$
		\State improvement $\gets$ $\emptyset$
		\For{{\bf each} children c $\in$ currChildren}
			
			\State recSol $\gets$ SingleBalance(c, smallRT, imbal)
			
			\If{recSol $\neq$ $\emptyset$}
				\State recImbal $\gets$ $|$ TaskCost(bigRT $\setminus$ recSol) - TaskCost(smallRT $\cup$ recSol) $|$
				\If{recImbal < imbal}
					\State improvement $\gets$ recSol
					\State imbal $\gets$ recImbal
				\EndIf
			\EndIf

			\If{TaskCost(c) $\notin$ uniqueChildrenCosts}
				\State uniqueChildrenCosts $\gets$ uniqueChildrenCosts $\cup$ TaskCost(c)
				\State uniqueChildren $\gets$ uniqueChildren $\cup$ c
			\EndIf

		\EndFor

		\For{{\bf each} children c $\in$ uniqueChildren}
			\State currImbal $\gets$ $|$ TaskCost(bigRT $\setminus$ c) - TaskCost(smallRT $\cup$ c) $|$
			\If{currImbal $<$ imbal}
						\State imbal $\gets$ currImbal
						\State improvement $\gets$ c
				\EndIf
		\EndFor

		\State {\bf return} improvement

		\end{algorithmic}
\end{algorithm}

The {\em SingleBalance} algorithm (see Algorithm \ref{alg:balance_single}) is divided into two parts, the recursion loop (lines 9-22) and the current level search loop (lines 23-29). Since the nodes are searched on a bottom-up breadth-first fashion, the first loop is responsible for recurring the {\em SingleBalance} operation on each of $bigRT$ child nodes (lines 9-10). The stop-condition for this recursion is when an empty bigRT is passed to {\em SingleBalance}, thus returning an empty improvement.

If an improvement is found (lines 11-13) it is then set as the new current best improvement (lines 13-16). Finally, the second loop (lines 23-29) goes through the current level children attempting to find improvements (lines 24-28), after which, the best improvement is returned (line 30) if any was found, or an empty improvement if no solution was found (line 8). 

\begin{algorithm}[!h]
\footnotesize
		\caption{The {\em Balance} step of the TRTMA\label{alg:balance}}

		\begin{algorithmic}[1]
		\State {\bf Input:} bucketList;
		\State{\bf Output:} bucketList;

		\State {\it bucketList is a sorted data structure by descending cost (e.g., multiset)}
		\While{true}
				\State bigRT $\gets$ bucketList.first()
				\State smallRT $\gets$ selectSmallRT(bucketList)
				\State imbal $\gets$ TaskCost(bigRT) - TaskCost(smallRT)
				\State improvement $\gets$ SingleBalance(bigRT.children, bigRT, smallRT, imbal)
				\State newMksp $\gets$ Max(TaskCost(bigRT $\setminus$ improvement), TaskCost(smallRT $\cup$ improvement))
				\If{improvement $\neq \emptyset$ {\bf and} newMksp $<$ TaskCost(bigRT)}
						\State bucketList $\gets$ bucketList $\setminus$ smallRT
						\State bucketList $\gets$ bucketList $\setminus$ bigRT
						\State smallRT $\gets$ smallRT $\cup$ improvement
						\State bigRT $\gets$ bigRT $\setminus$ improvement
						\State bucketList $\gets$ bucketList $\cup$ smallRT
						\State bucketList $\gets$ bucketList $\cup$ bigRT
				\Else
						\State break
				\EndIf
		\EndWhile
		\State {\bf return} bucketList

		\end{algorithmic}
\end{algorithm}

The {\em Balance} algorithm (see Algorithm \ref{alg:balance}) is implemented by repeatedly attempting to find an improvement from {\em SingleBalance} (line 8) until either an invalid improvement is returned (false balancing) or an empty improvement is returned (line 10). If any of those conditions apply then the {\em Balance} algorithm ends and returns the current state of $bucketList$ (line 21). It is worth noting that the $bucketList$ input must be a non-ascending ordered data structure, being this algorithm implemented with a C++ $multiset$ container with the task cost of each bucket as their keys. The reasons for choosing this structure is threefold: (i) it is sorted on insertions (with insertion $\mathcal{O}(log(n))$), (ii) it has a direct access operation on the beginning and end of the list ($\mathcal{O}(1)$), and (iii) the keys are not unique.

The first step of {\em Balance} is to select $bigRT$ and $smallRT$ (Algorithm \ref{alg:balance}, lines 5-6). The selection of $bigRT$ is done by taking the first bucket of $bucketList$. For $smallRT$ we can either select the last bucket of $bucketList$ (i.e., one of the buckets with the smallest task cost) or search among the buckets of $bucketList$ with the smallest cost for the one with the greatest reuse with $bigRT$. The later selection strategy can potentially result in more and better improvement opportunities available. The used approach was the last-bucket-selection with a full analysis on the impact of different $smallRT$ selection methods discussed further ahead.

After selecting $bigRT$ and $smallRT$, both are used on {\em SingleBalance} in order to search for an improvement on the current state of $bucketList$ (Algorithm \ref{alg:balance}, line 8). The returned improvement, if returned, is then validated (lines 9-10). If no improvement is returned, or if the returned improvement is a false balancing then the algorithm finishes its execution and returns the $bucketList$ as it currently is (lines 17-21). Otherwise, the improvement is applied to $bucketList$ (lines 11-16) and the algorithm searches for another improvement. Given the necessity for the ordering of $bucketList$, $bigRT$ and $smallRT$ are removed from $bucketList$ (lines 11-12), updated (lines 13-14) and then re-inserted on $bucketList$ on the right position (lines 15-16).

\paragraph{Algorithmic Complexity}

\begin{figure*}[b!]
\begin{center}
	\includegraphics[width=0.49\textwidth]{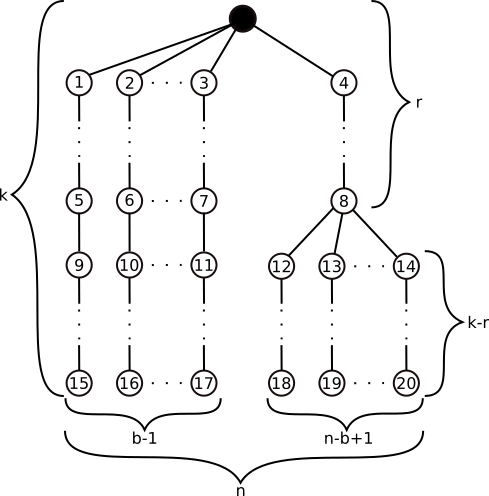}
	\caption{A general worst-case reuse-tree representation on which we have all $n$ stages divided into $b$ buckets. On this case we have $b-1$ buckets with exactly one stage, and thus cost $k$. Hence, the last bucket has $n-b+1$ stages. For this last bucket we assume the single and uniform reuse of the first $r$ task, having no reuse for the remaining $k-r$ tasks. This is the worst-case for balancing applications.}
\label{fig:complex}
\end{center}
\vspace{-4mm}
\end{figure*}

In order to calculate the computational complexity of the TRTMA we must first define a worst-case scenario on which the number of improvement attempts is maximum. One of this cases is defined in Figure \ref{fig:complex} with $\mathcal{O}(n)$ maximum improvement operations. This is the case on which $n/2-1$ of the $n/2$ buckets start with exactly one stage, and a single remaining bucket starts with $n-b+1$, with $b=MaxBuckets$. On this situation $n-b-1$ stages of the last bucket will be sent to another bucket on {\em SingleBalance} operations. Assuming that {\em SingleBalance} balances every pair of buckets with the minimum impact (improvements of exactly one stage), it will take $n-b-1$ balancing operations for all buckets to reach the final stable state of two stages per bucket. Thus, $\mathcal{O}(n)$ improvement operations.




For each improvement operation there is a selection step for $bigRT$ and $smallRT$, their update, and a {\em SingleBalance} call. The selection is done in $\mathcal{O}(1)$ for both subtrees since we are accessing the first and last elements of $bucketList$ (Algorithm \ref{alg:balance} lines 5-6). The update, which is comprised of two removal operations and two insertion operations are done in $\mathcal{O}(log(n))$ since $bucketList$ is an ordered data structure based on trees (Algorithm \ref{alg:balance} lines 11-16). For the {\em SingleBalance} call, exactly one balancing attempt is done for each traversed node on the worst case. Since for a graph with height $k$ and $n$ leaf nodes the number of nodes is bounded by $\mathcal{O}(kn)$, we have a final complexity of $\mathcal{O}(n$ $log(n) + kn^2)$. Also, given that $n \gg k$ the time complexity will be dominated by $\mathcal{O}(n^2)$.

\paragraph{Optimizations}

It is possible to reduce the cost of {\em SingleBalance} through two optimizations, already implemented on Algorithm \ref{alg:balance_single}: (i) single child pruning and (ii) unique sibling selection.

If a reuse-tree node $rtn$ is being visited by {\em SingleBalance}, and $rtn$ has only a single child node $rtn'$, then the improvement operation for both $rtn$ and $rtn'$ are the same. As such, we can prune $rtn$ from the search by moving down the subtree until either a leaf node is reached or a reuse-tree node with more than one child is found. This is implemented on Algorithm \ref{alg:balance_single}, lines 3-5.

\begin{figure*}[t!]
\begin{center}
	\includegraphics[width=0.49\textwidth]{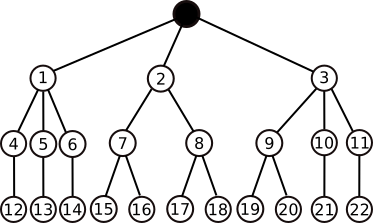}
	\caption{An example reuse-tree that can be used to illustrate possible prunable nodes. E.g., the use of nodes 4, 5, 6, or 10, 11 as an improvement attempt results in the same outcome (cost 3), making them interchangeable, as with nodes 7, 8 or 9 (cost 4), or nodes 12-22 (cost 3).}
\label{fig:prune}
\end{center}
\vspace{-4mm}
\end{figure*}

Furthermore, it is noticeable that any leaf node on the interval of S4-S9 of Figure \ref{fig:balance5} would result in the same balancing outcome (an imbalance of 0 with all buckets with cost 8). As such, it would be interesting if we pruned all nodes that would result in the same outcome. This can be, and is, achieved by verifying both the number of children and the cost of two nodes. If both values are the same than we have similar (or non-unique) nodes, meaning that only one of the nodes must be searched. This strategy is currently implemented locally, meaning that only sibling nodes are verified, which can be seen using Figure \ref{fig:prune}. This implementation is present on Algorithm \ref{alg:balance} on lines 18-21 and 23. For each child node traversed on {\em SingleBalance}, its task cost is calculated and, if it is unique (line 18), the matching child is added to a list of unique children (lines 19-20) to later be consumed (line 23).

By verifying prunable nodes locally it is meant that a node can only be pruned if the equivalent (repeated) search node is a sibling. On Figure \ref{fig:prune} this means that when searching the children of node 1, only node 4 would be further searched, being node 12 searched afterwards, ignoring nodes 5 and 6. As the search progresses, on the search of the children of node 2, only the nodes 7 and 15 would also be searched. Finally, nodes 9 and 10, and their children would be searched as well. However, by keeping a list of searched nodes, uniquely ordered by their children count and overall cost, it is possible to extend this strategy to a global scope, thus removing the sibling-only prunable node restriction. While using local prune on the reuse-tree of Figure \ref{fig:prune} would result in the search of 11 nodes (1, 4, 12, 2, 7, 15, 3, 9, 19, 10 and 21), a global prune scheme would result in 7 nodes searched (1, 4, 12, 2, 7, 15, 3).

In order to implement a global scope prune algorithm there is the need for both children count and overall cost metrics. Assuming that the reuse-tree of Figure \ref{fig:prune} does not have the subtree of node 3, both subtrees of nodes 1 and 2 would have the same overall cost (6). Thus, by considering only the overall cost, subtree 2 would not be searched, resulting in the missed opportunity of balancing with subtree 7 which has a cost of 4 (from the root node), an impossible value to achieve with only subtree 1 (which can achieve a costs 3 with nodes 1, 4 and 12, or 5 with nodes 1, 4, 12, 5 and 13). Likewise, by only verifying the children count on a reuse-tree with only the subtrees of nodes 1 and 3 we would come to the same fallacy of pruning a necessary subtree (this time, subtree of node 3), hence, making it necessary the use of both metrics.

\paragraph{Discussion on Additional Optimizations and Limitations}

\begin{figure}[b!]
	 \centering
	 \begin{subfigure}[t]{0.48\textwidth}
			 \centering
			 \includegraphics[width=0.5\textwidth]{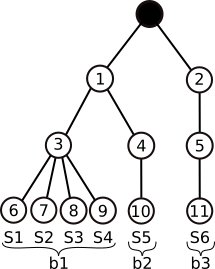}
			 \caption{Choosing the bucket with S5 results in a the premature finish of the TRTMA since there is not a single improvement between buckets b1 and b3.}
			 \label{fig:qual1}
	 \end{subfigure}
	 \hspace{1mm}
	 \begin{subfigure}[t]{0.48\textwidth}
			 \centering
			 \includegraphics[width=0.56\textwidth]{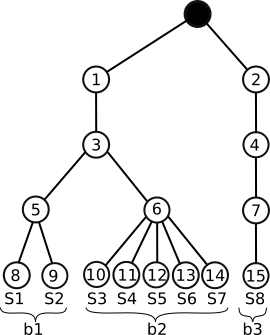}
			 \caption{Choosing to balance buckets b2 and b3 results in an imbalance of 1 with max cost 8 (imp = \{S7\}), while balancing b2 and b1 results in an imbalance of 0 with max cost 7 (imp = \{S7\}).}
			 \label{fig:qual2}
	 \end{subfigure}
	 \caption{Two examples of bad selection of $smallRT$ using the last-bucket strategy.}
	 \label{fig:qual}
\end{figure}

A limiting factor of the TRTMA is the $smallRT$ selection strategy. By trying an improvement with only a single $smallRT$ we may miss some better improvement opportunities, which may lead to better makespan values or even more balanced final results. The first possibility of improvement in the selection strategy arises from when two buckets of the same task cost can have different balancing outcomes when balancing with a given $bigRT$. This is exemplified on Figure \ref{fig:qual1}, where we have three buckets: $b1=\{S1,S2,S3\}$, $b2=\{S4\}$ and $b3=\{S5\}$, and either buckets $b2$ or $b3$ can be selected as $smallRT$ since they have the same cost, 3. If bucket $b3$ is selected then the TRTMA would finish prematurely since it does not exist an improvement between $b1$ and $b3$ that reduces the existing imbalance of 3 with max cost 6. However, for buckets $b1$ and $b3$ we have $imp={S3}$ which results in $b1=\{S1,S2\}$ and $b2=\{S3,S4\}$ with costs 5 and 5, thus showing a missed improvement opportunity.

This problem can be solved by selecting $smallRT$ as the bucket with the lowest task cost and also the highest reuse with $bigRT$. This solution was implemented and, across all tests, had negligible impact on the reuse attained by the TRTMA. Moreover, having to compare all $smallRT$ candidates with $bigRT$ has the execution time complexity $\mathcal{O}(n)$, since on the worst-case scenario we have $n/2-1$ buckets with one stage each (see Figure \ref{fig:complex}). Although the time complexity for TRTMA would not be changed, we would be increasing the reuse analysis execution cost to not achieve any benefits.

The second kind of missed improvements is shown in Figure \ref{fig:qual2}, on which the selection of $smallRT$ as one of the buckets with the smallest task cost (i.e., $b3$) results in missing the balancing of $smallRT = b1$, both with $bigRT = b2$. By attempting to balance $b2$ and $b3$ there exists no valid improvement. However, with $b2$ and $b3$ we have $imp=S7$, which results in buckets $b2$ and $b3$ with new cost 7 for both, improving the previous maximum task cost of 8.

In order to solve this problem the reuse between a single $bigRT$ and all remaining buckets would need to be calculated, which is basically an exhaustive search for all valid balancing and would have a combinatory-like time complexity. Preliminary testing has shown that the last-bucket selection strategy already achieves reuse degrees of close to 95\% of the reuse achieved by the RTMA for $MaxBucketSize = n$, for $n$ stages. As such, neither of these extra-reuse problems are worth being solved.

    \section{Experimental Results}

This chapter presents the experimental results of all proposed algorithms, regarding scalability, bucket cost balancing, the impact of different Sensitivity Analysis methods on reuse and the impact of the bucket size on run time.

\subsection{Experimental Environment}

We evaluated the proposed algorithms using a set of tissue images from brain cancer studies~\cite{kong2013machine}. The images were divided into 4K$\times$4K tiles for concurrent execution.  The image analysis workflow consisted of normalization, segmentation and comparison stages. The comparison stage computes the difference between masks generated and a reference mask set, created using the application default parameters. The experimental evaluations were conducted on two distributed memory machine environments. The first is the TACC Stampede cluster, with each node having dual socket Intel Xeon E5-2680 processors, an Intel Xeon Phi SE10P co-processor and 32GB RAM. The nodes are inter-connected via Mellanox FDR Infiniband switches. Stampede uses a Lustre file system accessible from all nodes. The second environment is the PSC Bridges cluster. Each node has a dual socket Intel E5-2695 and 128 GB RAM. Bridges uses a Pylon file system accessible from all nodes. The application and middleware codes were compiled using Intel Compiler 13.1 with ``-O3'' flag in both cases. All experiments were replicated at least 5 times and any claims for equivalence or difference between two algorithms of a given group were asserted through a t-test (two-tailed, not assuming homoscedasticity), on which $P < 0.001$ was chosen as the condition for assuming the difference to be statistically significant.

\subsection{Impact of Multi-level Computation Reuse for Multiple SA Methods} \label{sec:opt-search}

This section presents the impact of the computation reuse to the performance
of the MOAT and VBD SA methods. We first compute MOAT on all the application
parameters, because it demands a smaller per parameter sampling to exclude
those parameters that are non-influential to the output from the VBD. Most of
the experiments in this section were executed using a small number of machines,
because this section intended to detail the gains with the reuse optimizations.
However, Sections \ref{sec:bucket} and \ref{sec:scale} present experimental 
results for runs with large numbers of nodes.

\subsubsection{Impact of Multi-level Computation Reuse for MOAT}

\begin{figure}[h]
\begin{center}
	\includegraphics[width=1\textwidth]{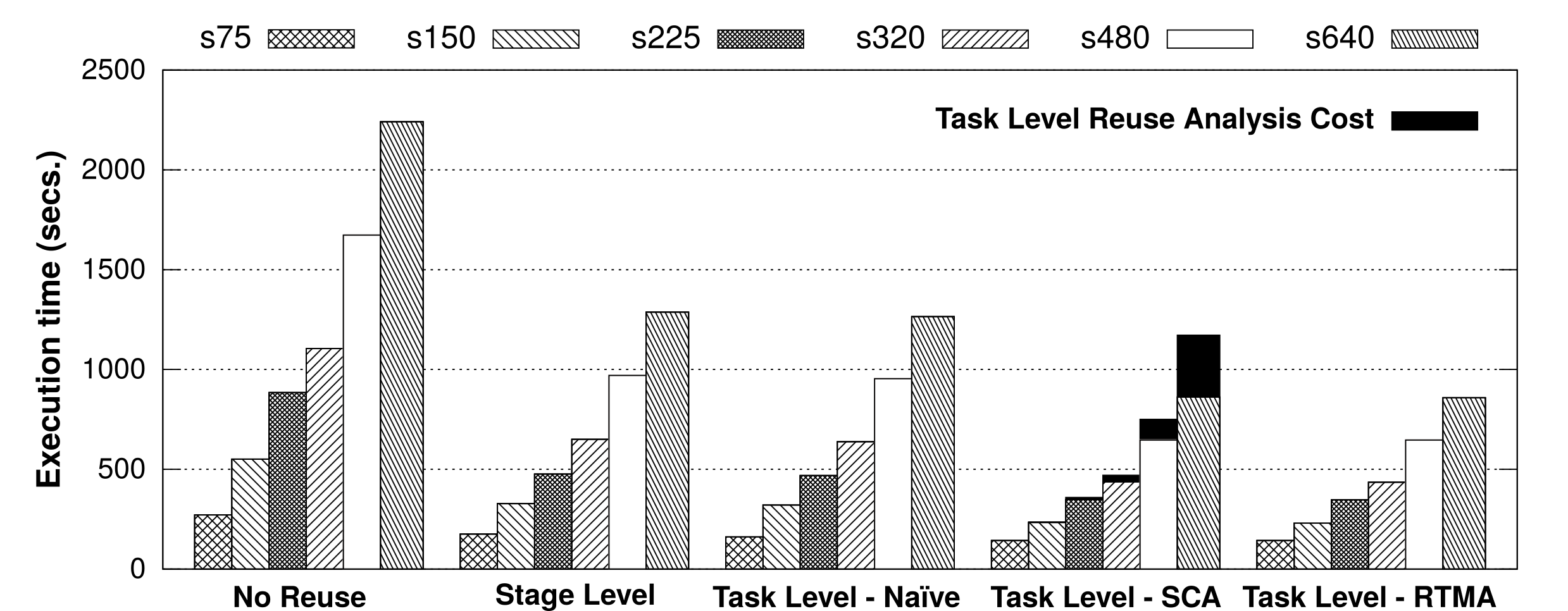}
	\caption{Impact of the computation reuse for different strategies as the sample size of the MOAT analysis is varied. E.g., s150 means that the experiment had 150 executions of the given workflow.}
	\label{fig:reuse-overall}
\end{center}
\end{figure}

Figure \ref{fig:reuse-overall} presents the execution times of MOAT studies
with parameter sample sizes varying from 160 to 640, which were executed using
only 6 Stampede nodes to demonstrate the impact of the optimizations. The parameters
were generated with a quasi-Monte Carlo sampling using a Halton sequence, which
is known to provide a good coverage of the parameter space. These experiments
use $MaxBucketSize$ set to 7, and the execution times refer to the makespan and
also include the cost to perform the computation reuse analysis and I/O.  
For the task level
merging approaches, the time spent by the merging algorithm is shown in the
upper part of the graph bars.  Additionally, five application versions were
executed: the ``No reuse'' that employs the replica based composition, the
``Stage level'' performs reuse only of stage instances, and the ``Task Level''
that reuses fine-grain tasks and is executed with the Na\"ive, SCA, and RTMA
algorithms. The TRTMA was not included on this analysis since for this scale it has the same performance as RTMA.

The results presented in Figure~\ref{fig:reuse-overall} show that all application versions that reused computation significantly outperformed the baseline ``No reuse'' version. The ``Stage Level'' reached a speedup of up to 1.85$\times$ on top of the ``No reuse'', while the application versions with ``Task Level'' reuse have higher gains. The ``Task Level - Na\"ive'' is only slightly better than the ``Stage Level'' (1.08$\times$ faster in the best case, being statistically distinct based on a t-test). This result is attributed to the highly order dependent nature of the na\"ive approach. The ``Task Level'' with SCA and RTMA, on the other hand, have remarkable speedups of up to, respectively, 1.39$\times$ and 1.5$\times$ on top of the ``Stage Level'' reuse only. 

It is also noticeable from Figure~\ref{fig:reuse-overall} that the performance gains with RTMA increase as the sample size grows and, as a consequence, more reuse opportunities are available. In the SCA algorithm, however, the opposite behavior is observed. This is a result of the higher costs of executing SCA to compute the stages to be merged, which offsets the gains with the actual execution of the application after the merging. The time taken by Na\"ive, SCA, and RTMA to compute the reuse are shown on the top of their bars on Figure~\ref{fig:reuse-overall}. For a sample of size 640, the time taken by SCA is about 26\% of the entire execution. It is also interesting to see that although the RTMA takes a much shorter time to compute the merging choices, it provides solutions as good as the ones returned by the SCA. In the best case, RTMA attained a speedup of up to 2.61$\times$ on top of the ``No reuse'' version. 

Regarding the atained reuse on the tested algorithms, both SCA and RTMA achieved values around 33\% of reuse. This value is the raw value of tasks that were not executed due to a merging algorithm. As such, the speedup of 1.5$\times$ of RTMA on top of ``Stage Level'' reuse, which is greater than the 33\% of reuse, is justified by the variable cost of each task. This means that the of 33\% of tasks that were not executed, or reused, were comprised of expensive tasks. A further analysis on the costs of tasks and the impact this variance has on the implemented approaches is present on Section \ref{sec:task-cost}

\subsubsection{Impact of Multi-level Computation Reuse for VBD}

\begin{figure}[t]
\begin{center}
	\includegraphics[width=1\textwidth]{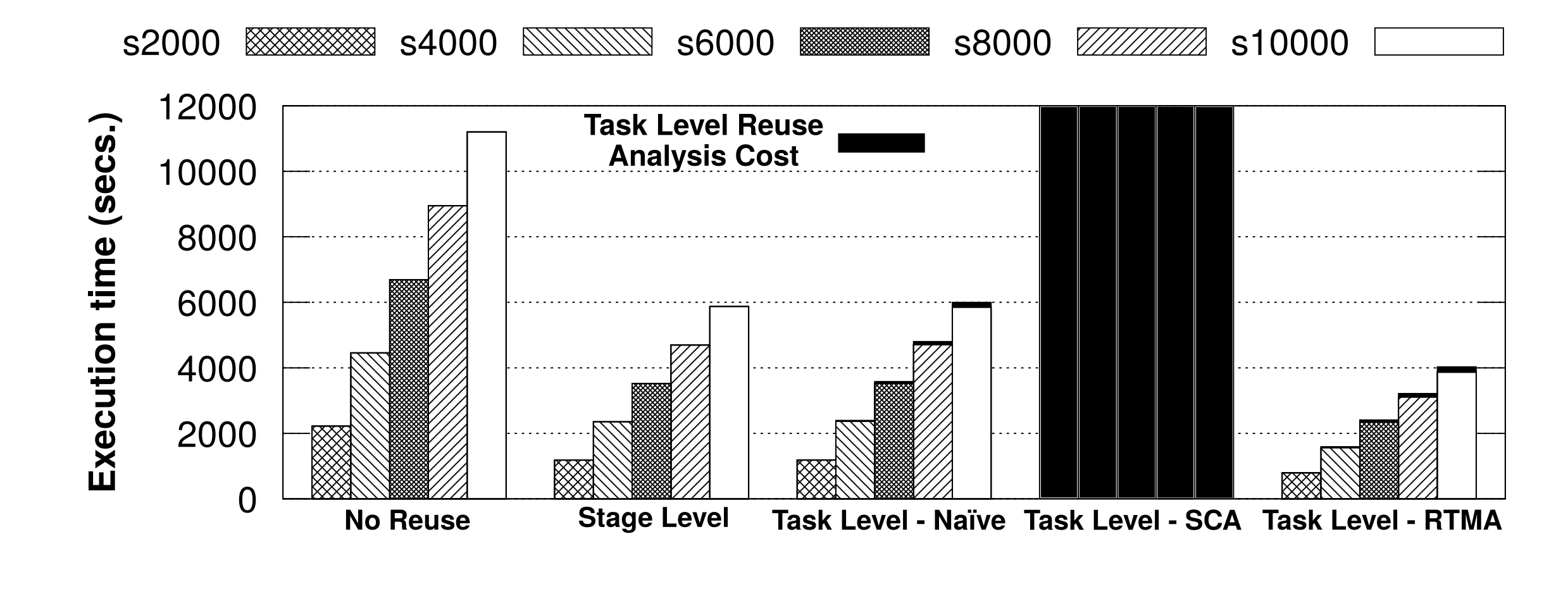}
	\caption{Impact of the computation reuse strategies for the VBD SA method.}
	\label{fig:vbd}
\end{center}
\vspace*{-3ex}
\end{figure}

The performance of the proposed optimizations for the VBD are presented in
Figure \ref{fig:vbd}. The VBD was executed using the 8 remaining parameters
(the original parameter set contains 15 parameters) that were not discarded in
the MOAT analysis. VBD requirements are of the order of hundreds to thousands runs per
parameter. As such, the sample size in this experiment is higher and was varied
from 2000 to 10000 runs, whereas the same application versions used with MOAT
were evaluated. In order to accelerate this analysis, we have increased the
number of nodes to 16 Stampede nodes.

As presented in Figure \ref{fig:vbd}, the relative performance of the application versions is similar to that observed with MOAT, except for the task level merging using SCA. Given that the sample size used in VBD is much higher, the SCA was not even able to finish computing the reuse to start up the actual execution of the workflow in 14000 secs. The RTMA had speedups of at most 2.9$\times$ against the ``No Reuse'' approach, and 1.51$\times$ on top of ``Stage Level''. These speedups were consistent with the ones found in the MOAT analysis. Similarly, the reuse for the VBD experiments was of at most 35\% for 10000 executions for the RTMA.

\subsection{SA Methods Reuse Analysis}
\label{sec:methods_results}

For all previous computation reuse tests which used the VBD method, the experiments were generated with the Latin Hypercube Sampler (LHS). Since the computation reuse on this work is highly reliant on the generated experiments, some sensitivity analysis methods were analyzed regarding their maximum reuse potential. Among them, in addition to LHS, the Monte-Carlo (MC) and Quasi-Monte-Carlo (QMC) methods were analyzed. The results are presented in Table \ref{tab:sa_methods}. This analysis is only performed for VBD given its continuous ranges of parameter values, which would present itself with less potential reuse when compared to MOAT methods and their discrete parameter value ranges.

\begin{center}
\begin{table*}[h]%
\centering
\begin{tabular*}{250pt}{@{\extracolsep\fill}lccc@{\extracolsep\fill}}
\toprule
Sample Size	&	200	&	600	&	1000 \\
\midrule
MC	&	36.35\%	&	36.46\%	&	36.40\% \\
LHS	&	36.62\%	&	36.44\%	&	36.44\% \\
QMC	&	35.10\%	&	34.44\%	&	33.48\% \\
\bottomrule
\end{tabular*}
\caption{Maximum computation reuse potential for MC, LHS and QMC methods with different sample sizes. For VBD, the number of experiments is $10 \times SampleSize$. The reuse percentages represent fine-grain reuse after coarse-grain reuse, meaning that only fine-grain reuse is being shown.\label{tab:sa_methods}}
\end{table*}
\vspace*{-3ex}
\end{center}

\subsection{Impact of Max Bucket Size}
\label{sec:bucket}

\begin{figure}[h!]
\begin{center}
	\includegraphics[width=0.9\textwidth]{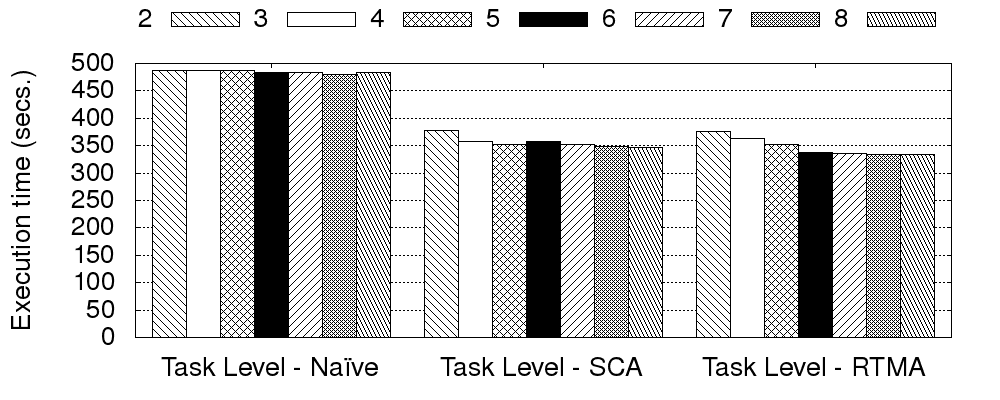}
	\caption{Impact of varying $MaxBucketSize$ from 2 to 8.}
	\label{fig:mbs}
\end{center}
\vspace*{-3ex}
\end{figure}

This section presents the impact of varying the $MaxBucketSize$ parameter on the execution times. As shown in Figure \ref{fig:mbs}, an increase in $MaxBucketSize$ leads to smaller execution times because of the larger number of merging opportunities. This increase has, however, a threshold, after which the maximum reuse for the experiment is achieved (usually arround 33\% of reuse, which results in speedups close to 1.5$\times$).

However, it interesting to notice that the variation in execution times as a result of the bucket size changes, when comparing the two ends ($MaxBucketSize$ 2 and 8), is up to 12\%, which shows that ``Task Level'' reuse can achieve significant gains even with small bucket sizes. This is result shows the viability of fine-grain reuse for execution environments on which there is a limited amount of memory available.

A large-scale SA experiment using the sample size of 240, 4,276 4K$\times$4K image tiles, and 128 Stampede computing nodes, using all optimizations and the ``No reuse'', ``Stage Level'', and ``Task Level RTMA'' versions of the workflow attained execution times of, 15,681s, 12,544s and 6,173s, respectively.

It is important to highlight that the task level merging reduces the number of stage
instances up to $MaxBucketSize$ times, and the parallelism as a consequence. This
could affect the application scalability if the number of stage instances after
the merging was not sufficient to completely use the parallel environment.


\subsection{The Effect of the Merging on Scalability}
\label{sec:scale}

\begin{figure}[h]
\begin{center}
    \includegraphics[width=1\textwidth]{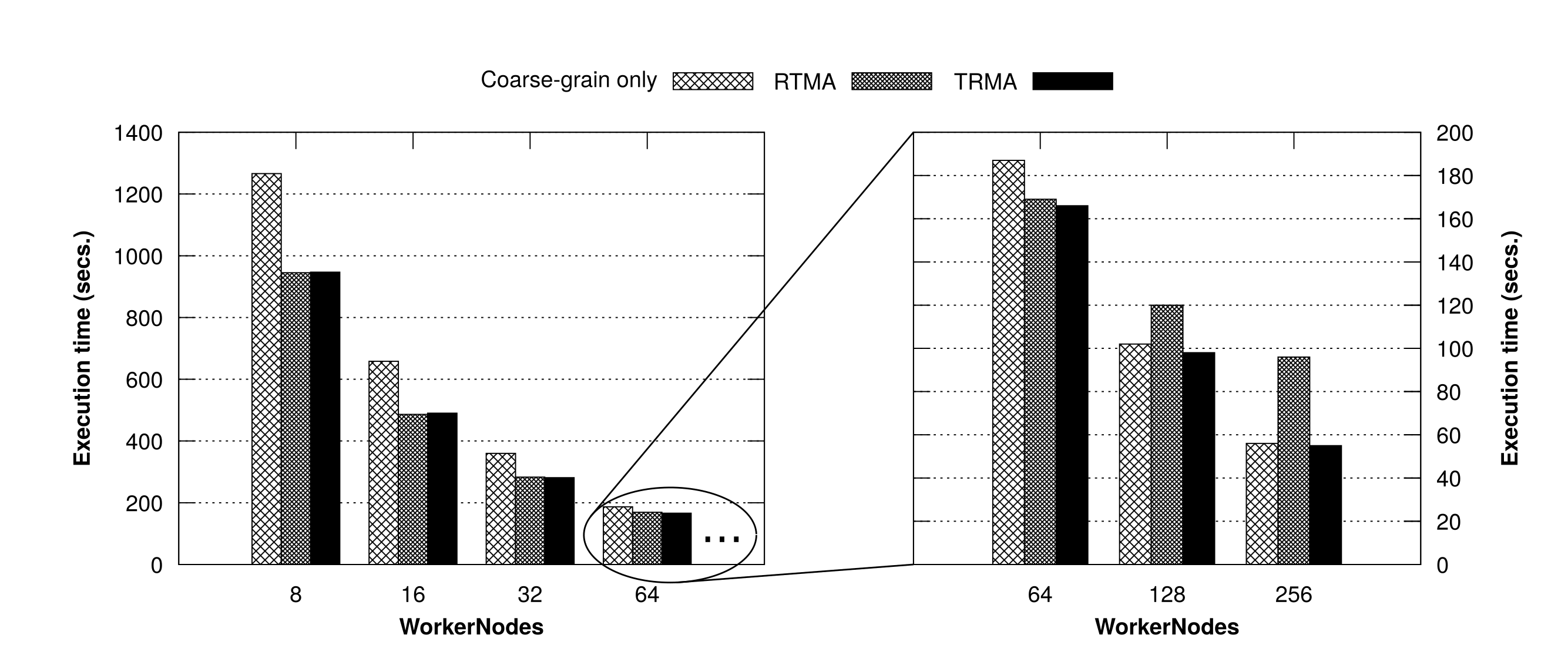}
	\caption{Comparison of the ``no fine-grain reuse'' (NR) approach with the RTMA and TRTMA. RTMA uses $MaxBucketSize$ 10, while TRTMA uses $MaxBuckets$ 3$\times$ the number Worker Processes (WP). The execution times for WP $>$ 32 were zoomed in a separated figure for the purpose of better visualization.}
	\label{fig:scale-new}
\end{center}
\vspace*{-3ex}
\end{figure}

This section evaluates the case on which performing merging operations may lead to poor scalability due to loss of parallelism. This problem is caused by the load imbalance of executing a different number of buckets on each node and can be triggered by either increasing the amount of merging performed or by increasing the number of nodes used. The later case was reproduced in Figure \ref{fig:scale-new} with the MOAT SA method and a sample size of 1000, with up to 256 Worker Processes/nodes (WP). 

\begin{figure}[h]
\begin{center}
    \includegraphics[width=1\textwidth]{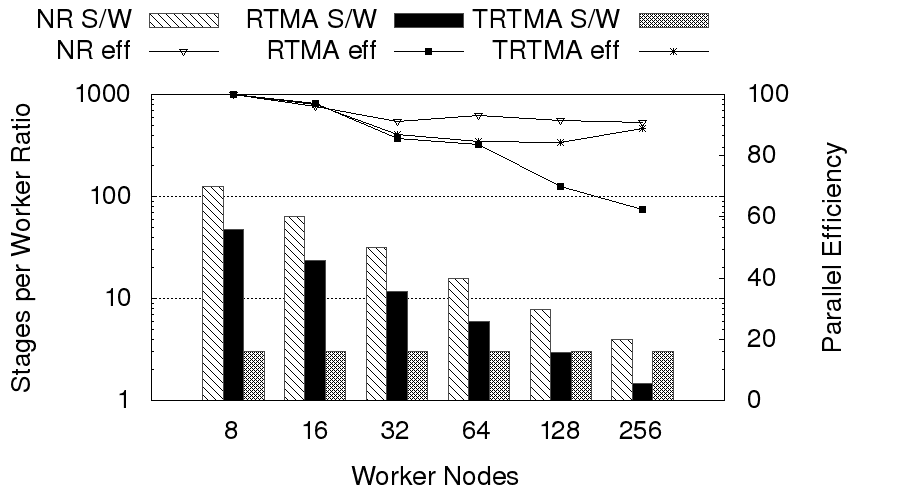}
	\caption{Combination of Stages per Worker Processes (S/W) and parallelism efficiency values. The S/W ratio for TRTMA was fixed as 3 for all WP values. The parallelism efficiency was calculated based on the previous execution (e.g. for WP 64, it is the execution time for WP 32 vs WP 64).}
	\label{fig:srs2}
\end{center}
\vspace*{-3ex}
\end{figure}

This performance degradation caused by excessive merging, as seen with the parallel efficiency of the RTMA on Figure \ref{fig:srs2}, is aggravated by the variable cost of different buckets generated by the RTMA. The workflow used on this work had its stages broken into finer-grain tasks in order to mitigate this variance on the costs. Since the RTMA generate buckets that are balanced stage-wise, but not task-wise, this difference in the number of tasks per bucket may lead to imbalance on environments with a low stages-per-worker ratio. This imbalance leads to a reduction of parallelism and, thereafter, degradation on the performance of the application due to load imbalance among nodes. On these cases the Task-Balanced Reuse-Tree Merging Algorithm (TRTMA) could be employed to extenuate this problem.

\begin{center}
\begin{table*}[b]%
\centering
\begin{tabular*}{460pt}{@{\extracolsep\fill}lcccccccc@{\extracolsep\fill}}
\toprule
Worker Processes (WP)&	8		&	16		&	32		&	64		&	128		&	256	\\
\midrule
Speedup TRTMA vs NR	&	1.33	&	1.34	&	1.27	&	1.12	&	1.04	&	1.01	\\
\midrule
TRTMA reuse			&	32.96\%	&	32.96\%	&	32.11\%	&	30.58\%	&	28.23\%	&	10.73\%	\\
\bottomrule
\end{tabular*}
\caption{Speedup of the TRTMA vs the ``No Reuse'' (NR) approach. \label{tab:speedup}}
\end{table*}
\vspace*{-3ex}
\end{center}

Still on Figure \ref{fig:srs2} it is visible that if the stages-per-worker ratio becomes low enough, the RTMA parallel efficiency drops to an extent on which it performs worse than not performing any fine-grain reuse at all. The values of stages-per-worker ratio (S/W), parallel efficiency and TRTMA reuse, compiled on Figure \ref{fig:srs2}, show that regardless the reuse algorithm employed, for the highest WP values the S/W ratio becomes low enough to impact the parallelism. This is true not only to RTMA, which becomes worse than ``No Reuse'' (NR) after WP 64, but also for NR itself. This loss of parallelism in NR is an indication of the imbalance between stages without reuse caused by the variance on the cost of tasks of the same level, but different inputs. Given that, the NR parallel efficiency values can be seen as the upper bound for any approach, since the reuse degree cannot increase for bigger WP values, nor can the parallel efficiency.

The TRTMA approach manages to improve on the RTMA parallel efficiency through bucket balancing, resulting in it not becoming worse than NR (see Figure \ref{fig:srs2} and Figure \ref{fig:scale-new}). The speedups that TRTMA achieves on top of NR lowers as WP increases, becoming negligible for WP values of at least 128 (see Table \ref{tab:speedup}). Given that for WP 256 the TRTMA attained 10.73\% of reuse, the speedup should either match this value or come close to it. This phenomenon of lack of performance is cause by another source of imbalance on buckets.

\subsubsection{The Impact of Variable Task Cost}
\label{sec:task-cost}

By taking another look at Figure \ref{fig:srs2} we can notice that the loss of parallelism due to imbalance starts at WP 32 for the RTMA and TRTMA approaches. This indicates that there exists another source of imbalance, for merging algorithms only, that affects RTMA harder than TRTMA and that is unaffected by TRTMA balancing techniques. It was found that this imbalance comes from the difference in the cost of tasks of different levels.

\begin{center}
\begin{table*}[b]%
\centering
\begin{tabular*}{460pt}{@{\extracolsep\fill}lcccccccc@{\extracolsep\fill}}
\toprule
Task 				&	t1	&	t2	&	t3	&	t4	&	t5	&	t6	&	t7	&	Total\\
\midrule
Avg Exec Time (s)	&	1.14	&	1.99	&	0.65	&	0.33	&	0.76	&	3.76	&	0.86	&	9.51	\\
Percentual			&	12.03\%	&	20.90\%	&	6.92\%	&	3.49\%	&	8.02\%	&	39.59\%	&	9.05\%	&	100\%	\\
\bottomrule
\end{tabular*}
\caption{An empirical evaluation on the costs of each task of which a stage is composed of. This approximation was generated with the purpose of showing the relative cost of the tasks, not being suitable as a absolute cost approximation. \label{tab:profile}}
\end{table*}
\vspace*{-3ex}
\end{center}

\begin{figure}[t!]
   \centering
   \begin{subfigure}[t]{0.9\textwidth}
       \centering
       \includegraphics[width=0.7\textwidth]{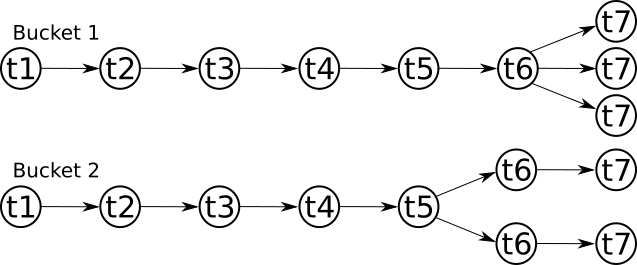}
       \caption{Two example buckets and their reuse trees. Bucket 1 was the result of the merger of three stages, while Bucket 2 had two stages initially.}
       \label{fig:unbal}
   \end{subfigure}
   \par\bigskip
   \begin{subfigure}[t]{0.9\textwidth}
   	   \centering
	   \begin{tabular*}{400pt}{@{\extracolsep\fill}lcccccccc@{\extracolsep\fill}}
			\toprule
			Task 				&	t1	&	t2	&	t3	&	t4	&	t5	&	t6	&	t7	&	Total\\
			\midrule
			Bucket 1	&	0.12	&	0.20	&	0.06	&	0.03	&	0.08	&	0.39	&	0.27	&	1.18	\\
			Bucket 2	&	0.12	&	0.20	&	0.06	&	0.03	&	0.08	&	0.79	&	0.18	&	1.48	\\
			\bottomrule
	   \end{tabular*}
		\caption{Sum of relative costs of tasks for each bucket. For a bucket containing only a single stage, and thus 7 tasks, the total cost would be 1.}
		\label{fig:unbal-tab}
	\end{subfigure}
	   \caption{An example case on which two buckets with the same number of tasks have different execution costs. This is due to the difference in the cost of different tasks. In this example Bucket 1 should execute 1.25$\times$ faster than Bucket 2.}
   \label{fig:unbal-ex}
\end{figure}

As shown in Table \ref{tab:profile}, the costs of the task which compose a stage are not constant. As such, buckets which are balanced by the number of tasks may still be susceptible to imbalance. An example of such case is presented in Figure \ref{fig:unbal-ex}. There, we have two buckets with the same number of tasks, but with different topologies. The first bucket was generated with three stages that attained maximum reuse, while the second had only two stages with less reuse. By using the TRTMA, the difference of execution cost between them of around 25\% would go unnoticed. This imbalance is enough to impact the parallel efficiency of an application through load imbalance. Effectively, this problem just makes the imbalance of buckets by tasks visible on an earlier S/W ratio.

Altogether, three sources of imbalance affects the maximum achievable parallel efficiency: (i) differently sized buckets (same stage count but different task count), (ii) buckets with the same size (task count) but different topologies, and (iii) same tasks having variant execution costs, which happens if two stages with the same topology and task count can have significantly different costs. The (i) problem is already solved by the TRTMA, while (ii) and (iii) can only be solved if we have an approximation of the costs of each task {\it a priori}.

    \section{Conclusion}

This work has proposed new algorithms that optimize Sensitivity Analysis (SA) through multi-level computation reuse. These algorithms were employed to optimize SA on a medical imaging analysis workflow, executed on large scale computation environments. Three fine-grain computation reuse algorithms were implemented, along with optimizations in order to deal with balancing, level of parallelism available and memory constraints.

The application selected for evaluating the proposed optimizations was a microscopy image analysis workflow. This workflow was chosen given its relevance \cite{rtf1,rtf2,DBLP:journals/bmcbi/KurcQWWTCNLSF15,DBLP:journals/ijhpca/SaltzTPCKKK13}, having a large sample space (around 21 trillion parameter combinations). The workflow is comprised of three stages, with the most expensive operation (segmentation) being composed of seven finer-grain tasks. On this workflow distinct SA methods were applied (MOAT and VBD) with several experiment generation methods (Section \ref{sec:methods_results}). Also, these analysis were tested on a large scale environment, running the Region Templates Framework (RTF) with at most 256 worker processes.

The RTF received two main improvements. The first is a way to easily generate workflows compatible with the RTF. This was achieved by using a descriptor file for the definition of each stage of the workflow, with a GUI to build and compose workflows based on this descriptor. These workflow compositions are performed with the assist of the Taverna Workbench \cite{taverna}, which provides an easy way to generate workflows for application experts.

Although computation reuse was an already studied strategy to reduce computational cost (Section \ref{sec:reuse_intro}), it was different from what was proposed by this work. The referenced approaches would either need a training step to be executed before the main application, which would be rather inefficient for a large scale workload such as the one used on this work; or perform computation reuse through caching methods, which would be too expensive to be employed on large scale computation environments. As such, the algorithms proposed on this work fill these limitations by performing computation reuse, in a lightweight manner.

Computation reuse was implemented and evaluated in two levels, stage-level and task-level. Stage-level computation reuse, implemented with a coarse-grain merging algorithm, was already proposed on previous works \cite{rtf1,rtf2} and re-implemented in this work. Although it already reduced the overall runtime by a large factor, some other computation reuse opportunities were unachievable through coarse-grain merging. Therefore, task-level computation reuse, implemented with fine-grain merging algorithms, was employed. One important feature of the fine-grain merging algorithms was that they could be used on top of coarse-grain merging results, augmenting their performance.

Out of the three fine-grain merging algorithms proposed, implemented and evaluated the Reuse-Tree Merging Algorithm (RTMA, Section \ref{sec:rtma}) stood out as an efficient approach. The RTMA achieved both high reuse factor (around 35\%) and low execution cost, when compared with the remaining approaches. 


It was identified that task balancing could be a problem if the ratio of tasks per core was low. In order to solve this problem a new approach based on the RTMA was implemented. This new approach, the Task-Balanced Reuse Tree Algorithm (TRTMA), was implemented to behave as the RTMA if the raw number of tasks is large enough that maximum parallelism is achieved, while also not degrading its performance if the tasks-per-core ratio was low. Moreover, the TRTMA was implemented with the intent to take only into consideration parallelism issues, by adjusting the $MaxBuckets$ parameter, which can be automatically chosen on runtime to optimize the application makespan while also taking the memory restrictions into consideration, thus reducing the dependency on the end user.


All algorithms were tested at first with the MOAT and VBD SA methods in order to assert their performance on real-world applications. It was shown that even though coarse-grain merging already had great speedups (from 1.85$\times$ to 1.9$\times$), fine-grain reuse managed to improve this values, achieving aggregate speedups between 1.39X to 1.51X on top of coarse-grain merging results, amounting to speedups of up to 2.89X. However, it is worth noting that the Smart Cut Algorithm (SCA) execution cost did not scale well, making this approach unfeasible for large scale setups.

The impact of the $MaxBucketSize$ constraint on the performance of the application was also analyzed, proving that the RTMA can be employed on heavily memory-constrained environments while also achieving good speedups. Since the TRTMA algorithm was equivalent to the RTMA on regular, large scale setups, only the worst-case scenario was tested. It was shown that even on this case, the TRTMA would always follow the best-case behavior. Finally, in order to validate the existence of computation reuse opportunities in the use case applications, and therefore validate the use of the proposed algorithms as a way to improve the makespan said applications, different SA experiment generators were tested in order to verify their maximum reuse degree. It was shown that across all cases the reuse degree was high enough to justify the use of computation reuse algorithms.

As a future work other application workflows would be to studied. For those applications the extensibility and ease of generating a new workflow from scratch would be observed. Then, it would be interesting to see the impact on reuse of differently structured workflows.

Another way to further optimize the workflow execution time through computation reuse is to perform balancing of buckets not by task count, but using the actual tasks costs. This approach would yield the best result, since there is not be any other source of loss of parallelism through the imbalance of buckets. However, cost analyzing is a difficult task which requires instrumentation and monitoring of such tasks \cite{instr1, instr2, instr3}, returning an estimation of these tasks costs. As such, the performance of this task-cost balancing can only be as good as the estimative of the tasks costs.

Furthermore, by balancing the buckets by task cost the bucket sizes could be limited only by parallelism and memory restrictions. The parallelism limitation is trivial to implement, being the number of buckets at least the number of worker processes. Again, the estimation is where the difficulty lies, being this memory consumption value rather hard to be found through static analysis \cite{mem_cost}. However, a task-cost balanced, memory-limited algorithm would attain not only maximum reuse, but maximum theoretical speedup since all limitations of computation would be solved.

    \bibliography{bibliografia}
    \bibliographystyle{ieeetr}

\end{document}